\documentclass{aa}
\usepackage{natbib}
\bibpunct{(}{)}{;}{a}{}{,} 
\usepackage{graphics}   
\usepackage{graphicx}
\usepackage{amsmath}
\usepackage{amsfonts}   
\usepackage{amssymb}  
\usepackage{mathrsfs}
\usepackage{epsfig,bbm}
\usepackage{epsfig}
\usepackage{bm}
\usepackage{color}
\usepackage{enumerate}
\usepackage{longtable}
\usepackage{epstopdf}
\usepackage[varg]{txfonts}
\usepackage{pdflscape}
\usepackage{hyperref}   
\usepackage[flushleft]{threeparttable}

\hypersetup{
    bookmarks=true,         
    unicode=false,          
    colorlinks=true,       
    linkcolor=blue,          
    citecolor=blue,        
    filecolor=blue,      
    urlcolor=blue           
}

\begin{document} 

   \title{Effects of axions on Population III stars}

   \author{
   Arthur Choplin
          \inst{1},
         Alain Coc
          \inst{2},
           Georges Meynet
           \inst{1},
          Keith A. Olive
          \inst{3},
          Jean-Philippe Uzan
          \inst{4},
          \and
          Elisabeth Vangioni
          \inst{4}
                    }

 \authorrunning{Choplin et al.}

	\institute{Geneva Observatory, University of Geneva, Maillettes 51, CH-1290 Sauverny, Switzerland\\
			e-mail: arthur.choplin@unige.ch
	\and Centre de Sciences Nucl\'eaires et de Sciences de la
		Mati\`ere  (CSNSM), CNRS/IN2P3, Univ. Paris-Sud, Universit\'e Paris--Saclay, 
		B\^atiment 104, F--91405 Orsay Campus (France)
	\and William I. Fine Theoretical Physics Institute, School of Physics and Astronomy, 
              	University of Minnesota, Minneapolis, MN 55455 (USA)
	\and Institut d'Astrophysique de Paris,
              UMR-7095 du CNRS, Universit\'e Pierre et Marie Curie,
              98 bis bd Arago, 75014 Paris (France),\\
              Sorbonne Universit\'es, Institut Lagrange de Paris, 98 bis bd Arago, 75014 Paris (France).        	     
		}
   \date{Received / Accepted}

  \abstract
   {}
   {Following the renewed interest for axions as a dark matter component, we revisit the effects of energy loss by axion emission on the evolution of the first generation of stars. These stars with zero metallicity are supposed to be massive, more compact, and hotter than subsequent generations. It is hence important to extend previous studies restricted to solar metallicity stars.}
	{Our analysis first compares the evolution of solar metallicity 8, 10 and 12~M$_\odot$ stars to previous work. We then calculate the evolution of 8 zero metallicity stars with and without axion losses and with masses ranging from 20 to 150~M$_\odot$.
	}
	{For the solar metallicity models, we confirm the disappearance of the blue loop phase for a value of the axion-photon coupling, $g_{a\gamma}=10^{-10}~{\rm GeV}^{-1}$. Regarding the Pop. III stars, we show that for $g_{a\gamma}=10^{-10}~{\rm GeV}^{-1}$, their evolution is not much affected by axion losses except within the range of masses 80--130 ~M$_\odot$. Such stars show significant differences both in their tracks within the $T_c$--$\rho_c$ diagram and in their central composition (in particular $^{20}$Ne and $^{24}$Mg). We discuss the origin of these modifications from the stellar physics point of view, and their potential observational signatures.}	
	{}

   \maketitle

	\titlerunning{Effects of axions on Population III stellar evolution}
	\authorrunning{A. Choplin et al.}



\section{Introduction}

Two highly motivated and very well studied candidates for  dark matter are the lightest supersymmetric particle \citep{Goldberg83, Ellis84} and the axion \citep{Preskill83, Abbott83, Dine83}. While the former is under intense pressure due to null searches for supersymmetry at the LHC \citep{Bagnaschi15}, axions remain a viable dark matter candidate.  
Axions were originally proposed as a possible solution to the strong CP problem~\citep{Peccei77b, Peccei77a, Weinberg78, Wilczek78}. In strong interactions, there is no reason for CP violating effects to be small, and the presence of such interactions
would disagree violently with experiment unless the coefficient of the CP violating term, called $\theta$ is tuned to be very small
($< 10^{-10}$).  However, the spontaneous breaking of a global U(1) (Peccei-Quinn (PQ) symmetry) allows for the possibility of a dynamical cancellation of the CP violating phase in QCD. If the scale associated with the symmetry breaking, $f_a$, is large, interactions between the axion and matter become very weak rendering the  axion nearly invisible \citep{Kim79, Shifman80, Zhitnitskii80, Dine81}.  Because the PQ symmetry is also explicitly broken (the CP violating $\theta$ term is not PQ invariant) the axion picks up a small mass similar to the pion picking up a mass when chiral symmetry is broken.  Roughly, $m_a  \sim m_\pi f_\pi /f_a$  where $f_\pi \approx 92$ MeV, is the pion decay constant, so that
\begin{equation}\label{e.ma}
m_a \approx (6 \times 10^6 {\rm GeV}/f_a)\,{\rm eV}.
\end{equation}

 As dark matter candidates, despite their low mass (if $f_a \gg f_\pi$), axions act as cold dark matter as the cosmological energy density in axions consists of their coherent scalar field oscillations.  The energy density stored in the oscillations exceeds the critical density~\citep{Preskill83, Abbott83, Dine83} unless $f_a\lesssim 10^{12}$~GeV or $m_a \gtrsim 6 \times 10^{-6}$ eV.

Although model dependent, the axion has couplings to photons and matter fermions. As a result, they may also be emitted by stars and supernovae \citep{Raffelt90}. In supernovae, axions are produced via nucleon-nucleon bremsstrahlung with a coupling $g_{AN} \propto m_N/f_a$.  SN 1987A enables one to place an upper limit~\citep{Ellis87, Mayle88, Mayle89, Raffelt88, Raffelt91, Burrows90, Keil97} on the axion mass of 
\begin{equation}\label{malim}
m_a  \lesssim(0.5 - 6) \times 10^{-3}\,{\rm eV}.
\end{equation}
Axion emission from red giants implies~\citep{Dearborn86, Raffelt95}   $ m_a  \lesssim 0.02$~eV (though this limit depends on the model dependent axion-electron coupling).  
From Eq.~(\ref{e.ma}), the limit in Eq.~(\ref{malim}) translates into a limit 
$f_a \gtrsim (1-12) \times 10^{9}$
implying that only a narrow window exists for axion masses.

In most models, the axion will also couple electromagnetically to photons through the interaction term $\mathcal{L} = - g_{a\gamma} \phi_a \bf{E\cdot B}$, where $\phi_a$ is the axion field \citep{Raffelt90}. While not competitive in terms of axion mass limits, the axion-photon coupling, $g_{a\gamma}$, can be constrained through several different stellar processes. In low mass stars, while axion emission occurs through the lifetime of the star,  emission at high temperatures can greatly reduce the lifetime of the helium burning phase resulting in an upper limit of $g_{10}{\equiv}g_{a\gamma}/(10^{-10}~{\rm GeV}^{-1}) < 0.66$~\citep{Raffelt87, Ayala14}. A similar limit of $g_{10} < 0.8$ was derived from the evolution  of more massive stars (in the range of 8-12 M$_\odot$) as energy losses in the helium
burning core would overly shorten the blue-loop phase of stellar evolution in these stars~\citep{Friedland13}. For recent reviews, see ~\cite{Kawasaki13} and \cite{Marsh16}.

Here we consider the effect of axion emission in more massive but metal-free stars associated 
with Pop. III. In such stars we can expect enhanced axions losses compared to solar metallicity stars for at least two reasons. First, one expects larger initial masses for Pop. III stars. The lack of heavy elements prevents dust to be formed and dust is a key agent at solar metallicity to fragment proto stellar clouds in small masses. This is the reason why very few low, long lived stars are believed to be formed in metal free environments. Only massive or even very massive stars are expected to form \citep[see for instance the review of][]{Bromm13}. The second reason is that the absence of heavy elements implies that stars of a given initial mass are more compact and thus reach higher central temperatures \citep[see the discussion in][for instance]{Ekstrom08}.

After describing the process of energy loss by axions in Sect.~\ref{sec2}, we confirm in Sect.~\ref{sec3} using the Geneva code of stellar evolution that axion emission from solar metallicity stars might be constrained by the disappearance of the blue-loop phase. 
Our goal, however, is to study the axion energy losses for massive Pop. III stars, which is described extensively in Sect.~\ref{sec4}.
Our conclusions are given in Sect. \ref{sec5}.

\section{Energy loss by axions}\label{sec2}

\begin{figure}[h]
\begin{center}
 \includegraphics[width=.445\textwidth]{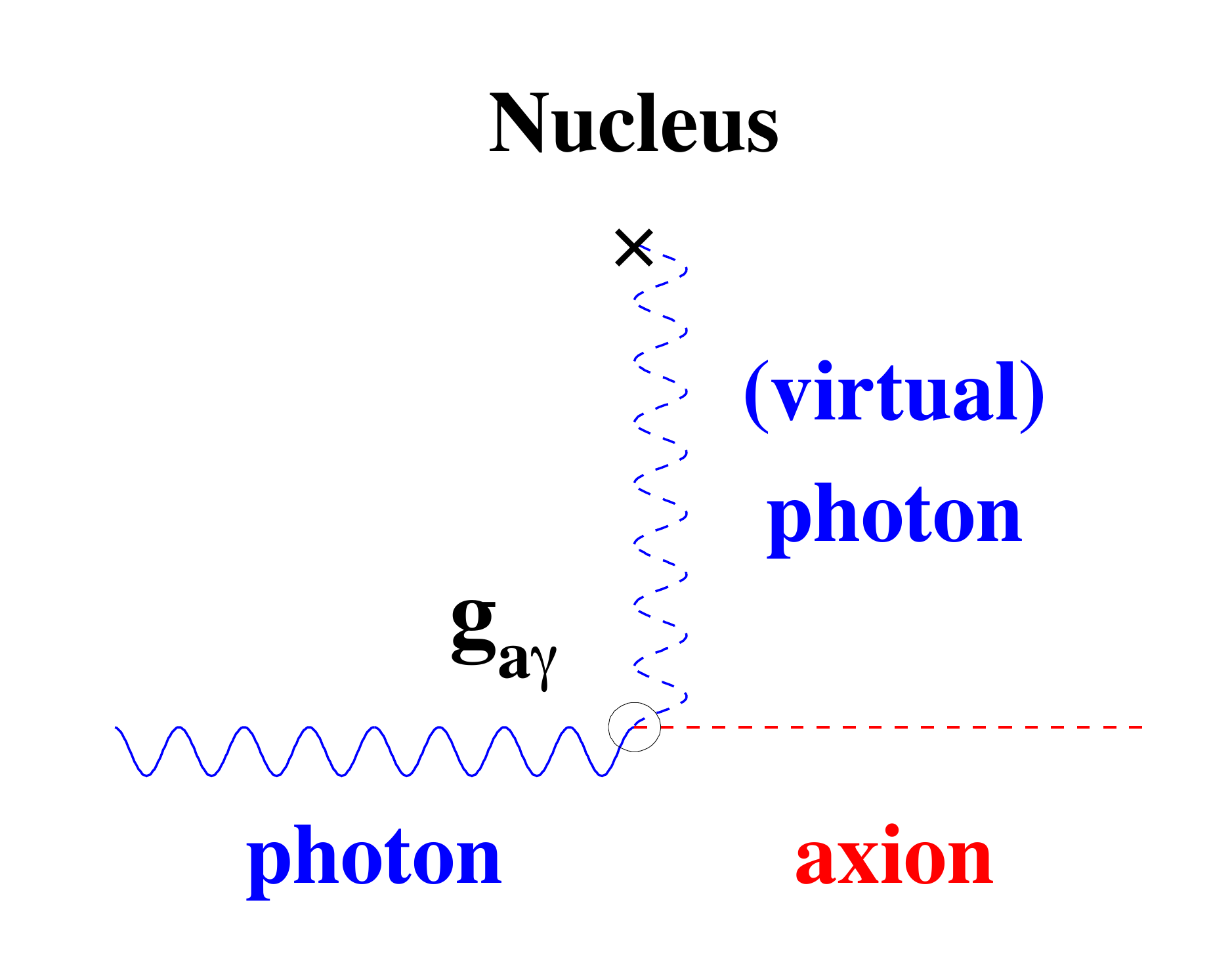} 
\caption{The Primakoff effect: a real photon from the thermal bath is converted to an axion by the electric field of a nucleus.}
\label{f:prima}
\end{center}
\end{figure}

For a given coupling of axions to photons, Primakoff emission \citep{Primakoff51} of axions from photon-nucleus scattering will occur, mediated by the
virtual photons from the electrostatic potential of the nucleus  (Fig.~\ref{f:prima}). The screening of this potential by the freely moving electric charges (Debye--H\"uckel effect), 
however, need to be taken into account \citep{Raffelt86}.
The volume emissivity for this process as a function of temperature (for temperatures much greater than the plasma frequency) was computed in~\cite{Raffelt86,Raffelt90}. 
Dividing this emissivity by the mass density, $\rho$ we obtain an energy loss rate per unit mass,
\begin{equation}
\epsilon_{\rm ax}=\frac{g_{a\gamma}^2T^7}{4\pi^2\rho}\xi^2f(\xi^2)
\label{q:loss}
\end{equation}
where the function $f$ is defined in Eq.~(4.79) of~\cite{Raffelt90} and $\xi$ is given by
\begin{equation}
\xi\equiv\frac{\hbar\mathrm{c}\;k_S}{2\mathrm{k_B}T}.
\label{q:xi}
\end{equation}
with $\hbar$ the reduced planck constant, $\rm c$ the speed of light and $\rm  k_B$ the Boltzmann constant (see the Appendix for explanation of units).
The Debye--H\"uckel screening wavenumber, $k_S$ is given by~\cite{Raffelt08}:
\begin{equation}
k_S^2\equiv4\pi\alpha\left(\frac{\hbar\mathrm{c}}{\mathrm{k_B}T}\right)\sum_{i=e,{\rm ions}}{n_i}Z_i^2
\label{q:debye}
\end{equation}
where $\alpha$ is the fine structure constant, $Z_i$ the atomic number and $n_i$ the ions/electron number densities, given by
\begin{equation}
n_{\rm ions}=\rho\frac{X_i}{A_i}\mathcal{N}_A\;\mathrm{and}\;n_\mathrm{e}=\sum_{i={\rm ions}}{n_i}Z_i,
\end{equation}
(because of charge neutrality) with $\mathcal{N}_A$ the Avogadro number.
The energy loss rate (Eq.~\ref{q:loss}) can be rewritten as:
\begin{equation}
\epsilon_{\rm ax}=283.16\times g_{10}^2T_8^7\rho_3^{-1}\xi^2f(\xi^2)\;\mathrm{erg/g/s},
\label{q:lossx}
\end{equation}
where $T_8{\equiv}T/(10^8$~K) and $\rho_3{\equiv}\rho/(10^3 \mathrm{g/cm^3})$. Note that the numerical constant differs from the one in Eq.~(3) of~\cite{Friedland13} by one order of magnitude (see the Appendix for more details).  

We correct Eq.~(\ref{q:loss}) by the damping factor introduced by~\cite{Aoyama15}, namely, $\exp(-\hbar\omega_0/{\mathrm{k_B}T}$) where $\omega_0$ is the plasma frequency in MeV:
\begin{equation}
\hbar\omega_0=\left[4\pi\alpha\left(\frac{\hbar\mathrm{c}}{m_e c^2}\right)n_e\right]^\frac{1}{2}\hbar\mathrm{c}.
\label{q:damp}
\end{equation}
However, the damping factor stays always of order unity in our models.

In massive stars, during the Main-Sequence (MS), most of the energy is transported by radiation and convection so that the axions have little effect. Axion cooling is believed to have a significant effect during the core helium burning phase, when the central temperature and density are about $10^8$ K and $10^3$ g cm$^{-3}$ \citep[see e.g.][]{Friedland13}. In the next stages (core carbon burning, oxygen,...), because of the higher temperature and density, the cooling by axions would be more pronounced. At these late stages however, the axion cooling competes with neutrino losses. Whether axion losses, in Pop.III stars, are significant or not compared to other sinks of energy, like neutrino losses, is a point investigated in this paper.

\begin{figure}[htb]
\begin{center}
 \includegraphics[width=.485\textwidth]{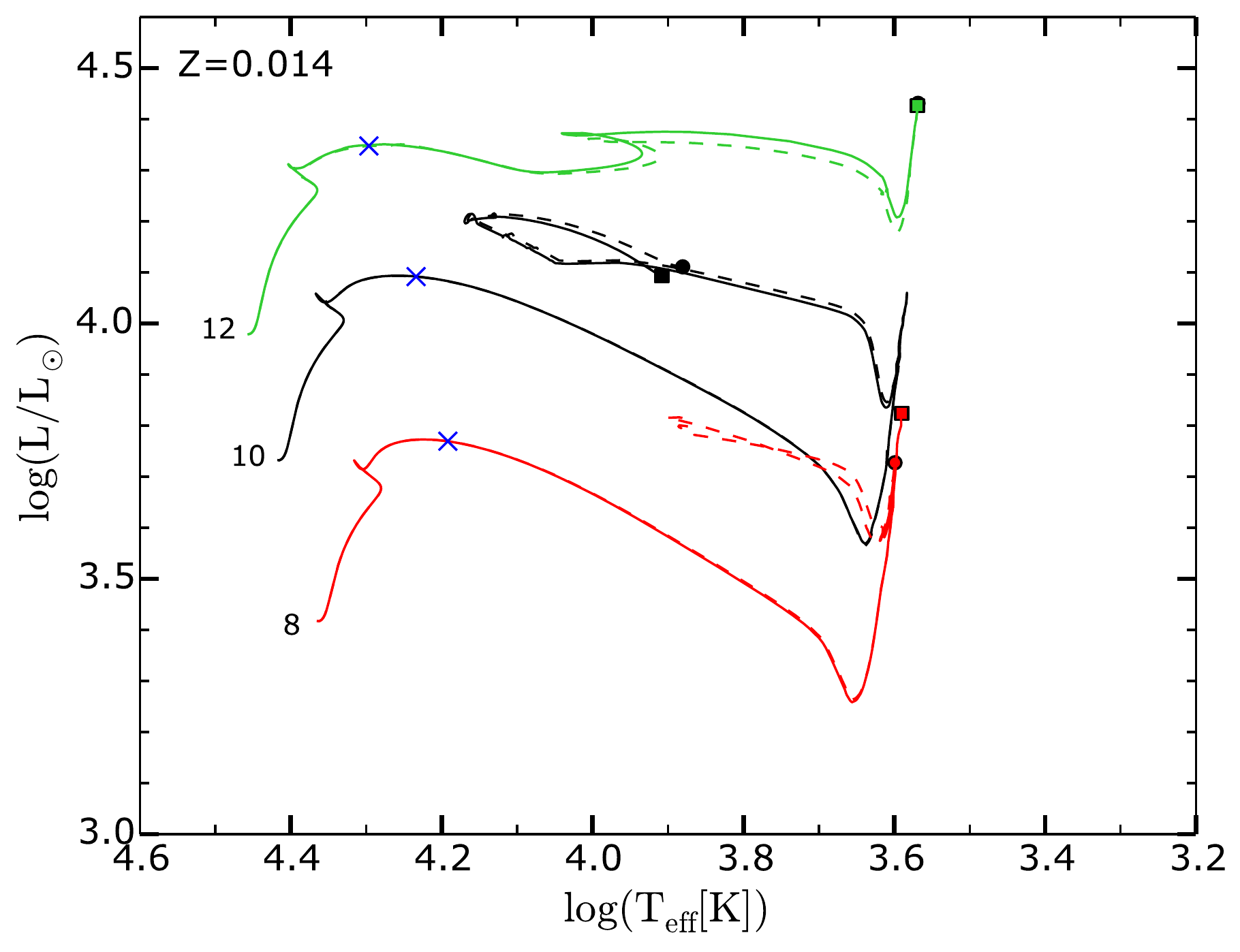} 
\caption{Hertzsprung-Russell diagram for the 6 models computed at solar metallicity. Dashed lines show models without axion losses. Solid lines show models with axion losses. Circles (without axion losses) and squares (with axion losses) show the location of the models at the end of core He-burning. Blue crosses show the location of the tracks at core helium burning ignition ($\epsilon$($3\alpha$) $>0$).}
\label{f:ht}
\end{center}
\end{figure}

\section{Massive stars with solar metallicity}\label{sec3}

It was shown that at solar metallicity the energy losses by axions from the helium-burning core can eliminate the blue loop phase~\citep{Friedland13}. Limits on the axion photon coupling were derived from studies of  $8-12 M_{\odot}$ models. To check whether we find similar results, we have computed models at similar mass and metallicity.

\subsection{Physical ingredients}

We computed 6 models of 8, 10 and 12 $M_{\odot}$ at solar metallicity, with and without energy losses by axions using the Geneva code.
Similar input parameters, as those used in~\cite{Ekstrom12}, have been considered for these computations. The initial abundance of H, He and metals, in mass fraction, are $X = 0.72$, $Y=0.266$ and $Z=0.014$ respectively. 
The mixture of heavy elements is determined according to \cite{Asplund05} except for Ne, whose abundance is taken from \cite{Cunha06}. The isotopic ratios are from \cite{Lodders03}.
The core overshoot parameter $d_{\rm over}/H_{\rm p} = 0.1$, where $H_p$ is the pressure scale height determined at the Schwarzschild convective core boundary. It extends the radius of the convective core by an amount of 0.1 $H_p$. The outer layers, if convective, are treated using the mixing length theory. The mixing-length parameter $\alpha_{\rm MLT}$ is set to 1.6. This value allows one to reproduce the solar radius at the age of the sun and the positions of the red giants and supergiants in the HR diagram \citep{Ekstrom12}.

To compute the energy loss rate per unit mass due to axions,  we have taken $g_{10} = g_{a\gamma}/(10^{-10}~{\rm GeV}^{-1}) = 1$ in all our models. The simulations are stopped at the end of the core helium burning phase. The model are labelled as `10g1' for instance, where `10' refers to the initial mass in solar masses and `g1' means $g_{10}=1$.

\subsection{Evolution in the Hertzsprung Russell diagram}

Fig. \ref{f:ht} depicts the evolutionary tracks in the HR diagram. For the 8 $M_{\odot}$ models (red lines), we see that when the energy loss by axions is included (solid red line), the blue loop disappears. To understand this behavior, let us recall that a  blue loop appears when the stellar envelope contracts causing the star to leave the red supergiant branch and to evolve blueward in the HR diagram. During the core He-burning phase, the envelope contracts due to the core expansion. This is due to the mirror effect. The expansion of the core comes from the fact that when the abundance of helium decreases in the central regions, the central temperature increases. An effect of the increase of the central temperature is the increase of the nuclear energy generation by reactions like for instance $^{12}$C($\alpha$,$\gamma$)$^{16}$O (we note here that the rate of this reaction is still highly uncertain). The excess of nuclear energy is used to expand the core. 
The appearance/disappearance of the blue loops is very sensitive to many inputs of the stellar models and small changes can have important effects, as changes of the mesh resolution, the way to account for convection, the mixing in the radiative zones \citep[see e.g.][]{Kippenhahn90, Maeder01, Walmswell15}.

When energy losses by axions are included, the energy produced by the nuclear reactions in the core is more efficiently removed from the helium burning core (the energy removed by axions is $\lesssim 3 \times 10^4$ erg g$^{-1}$ s$^{-1}$ for the models of this section). This limits the expansion of the helium core, that in turn limits the contraction of the envelope. This finally tends to reduce or even prevent the formation of a blue loop. This change of behavior is restricted to the 8 M$_\odot$ star among the three different initial mass models considered here.

\subsection{Duration of burning phases}

Interestingly, for our 8 $M_{\odot}$ model, the duration of the helium burning phase is reduced by $\sim 13 \%$ when axions are considered. For the $10$ and $12$ $M_{\odot}$ models, it is $\sim 8 \%$ and $\sim 11 \%$ respectively.  Thus we see that the largest effects on the lifetimes are found for the 8 M$_\odot$ case. This does appear somewhat consistent with the fact that in this model the blue loop is affected. The lifetimes are shorter when axions are considered simply because they add a new channel that is very efficient for removing energy from the core. 

\cite{Friedland13} found a decrease of the duration of helium burning by $\sim 23 \%$ (cf. their Fig. 3 and discussion) for a $9.5$ $M_{\odot}$ with $g_{10} = 0.8$, thus larger than the lifetime decreases that we have obtained here. This might appear surprising in view of their value for  $g_{10}$ that is lower than the one adopted here. Instead, one would have expected that their lifetime decreases would be smaller than those obtained here with a $g_{10}=1$. In all likelihood, our stellar models differ by some other physical ingredients, such as the overshoot parameter. These differences are however difficult to trace back due to the lack of details given in~\cite{Friedland13}.  More importantly is the fact that we find similar qualitative trends. Using a different stellar evolution code, we find as in~\cite{Friedland13}, that axions  suppress the blue loops for some masses and decrease the core He-burning lifetime.

Contrary to the helium burning phase, the duration of the MS is very little affected by axions because the relevant temperature for axion cooling to be efficient is not reached in the core of massive MS stars. The duration of this phase is reduced by less than $\sim 0.1 \%$ for our models. This is because the losses by axions are very small during that phase.

\begin{figure}[t]
\begin{center}
 \includegraphics[width=.485\textwidth]{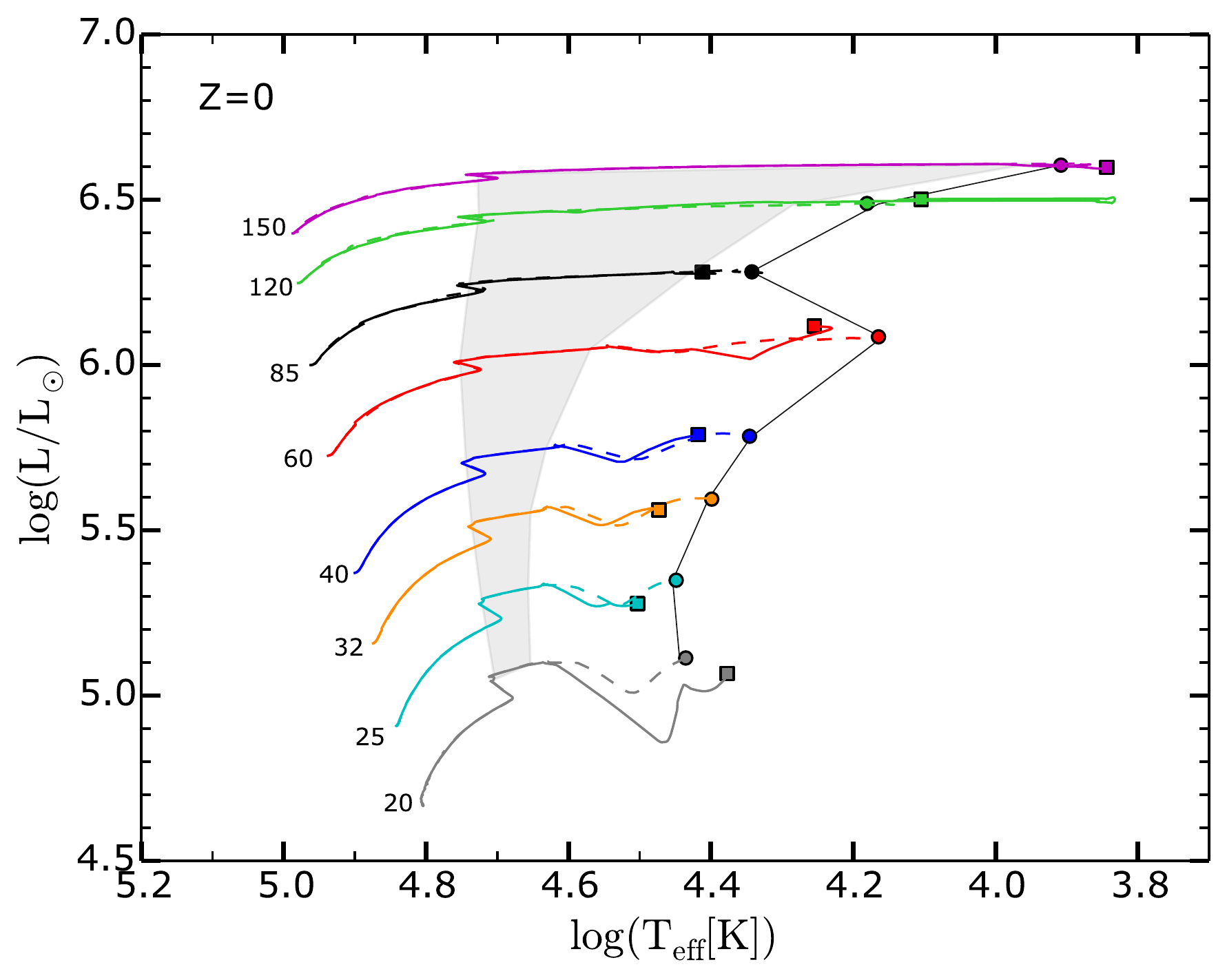} 
\caption{Same as Fig. \ref{f:ht} but for the 16 Pop. III models. The shaded area shows the zone of the diagram where helium burns in the core of the models without axion losses. The thin black line shows the location of the models without axion losses when the carbon starts to burn into the core. The shaded area and the black line are similar for models with axion losses. Circles (without axion losses) and squares (with axion losses) show the location of the models at the end of core C-burning.}
\label{f:htpop3}
\end{center}
\end{figure}

\section{Pop. III stars}\label{sec4}


\subsection{Physical ingredients}

\begin{figure}[t]
\begin{center}
 \includegraphics[width=.485\textwidth]{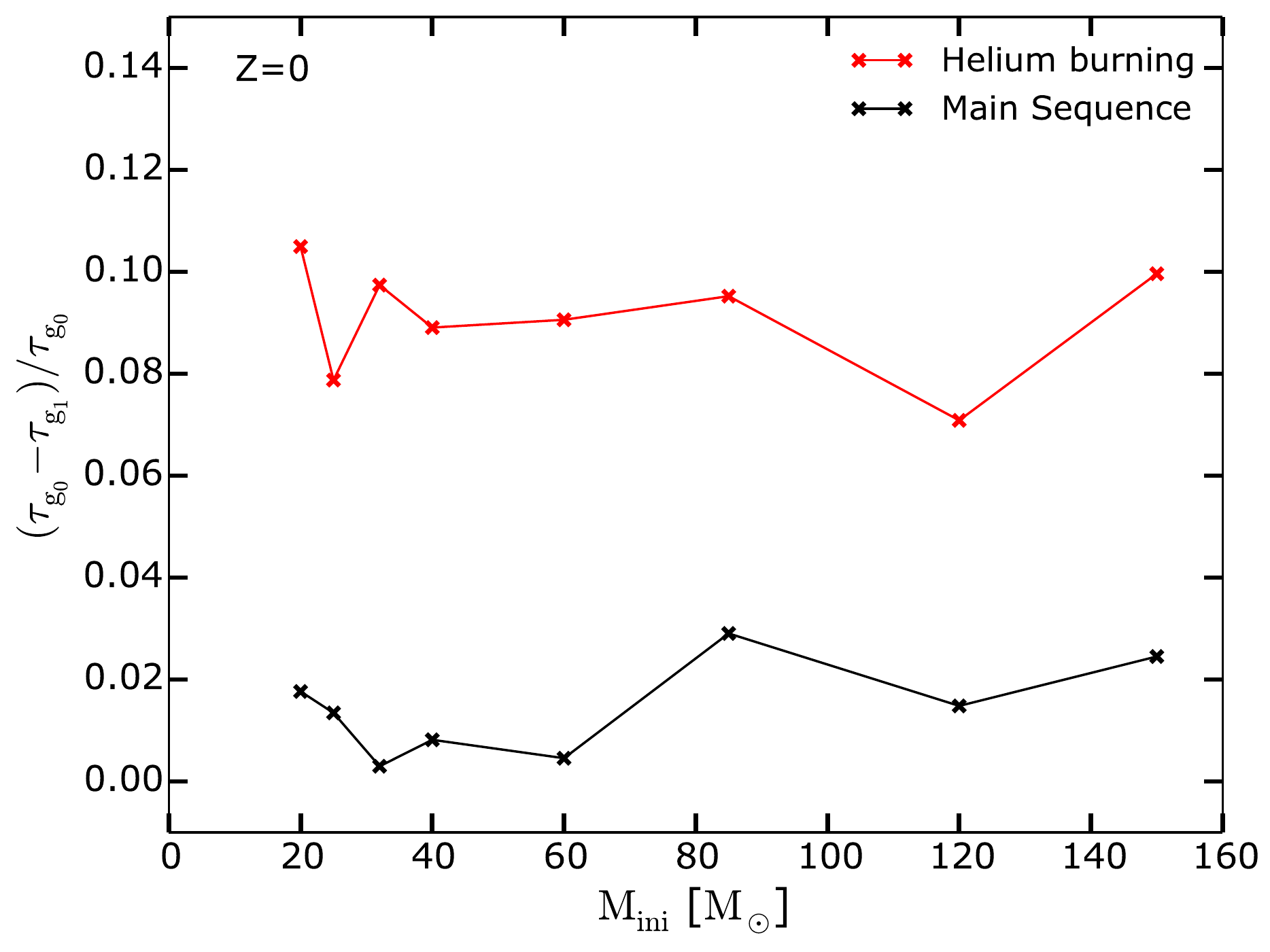} 
\caption{Relative difference of the duration of Main-Sequence and core helium burning phases between Pop. III models with ($g_{10} = 1$, noted as $g_1$) and without ($g_{10} = 0$, noted as $g_0$) axions losses.}
\label{f:mtau}
\end{center}
\end{figure}

\begin{figure*}[t]
\begin{center}
 \includegraphics[width=.98\textwidth]{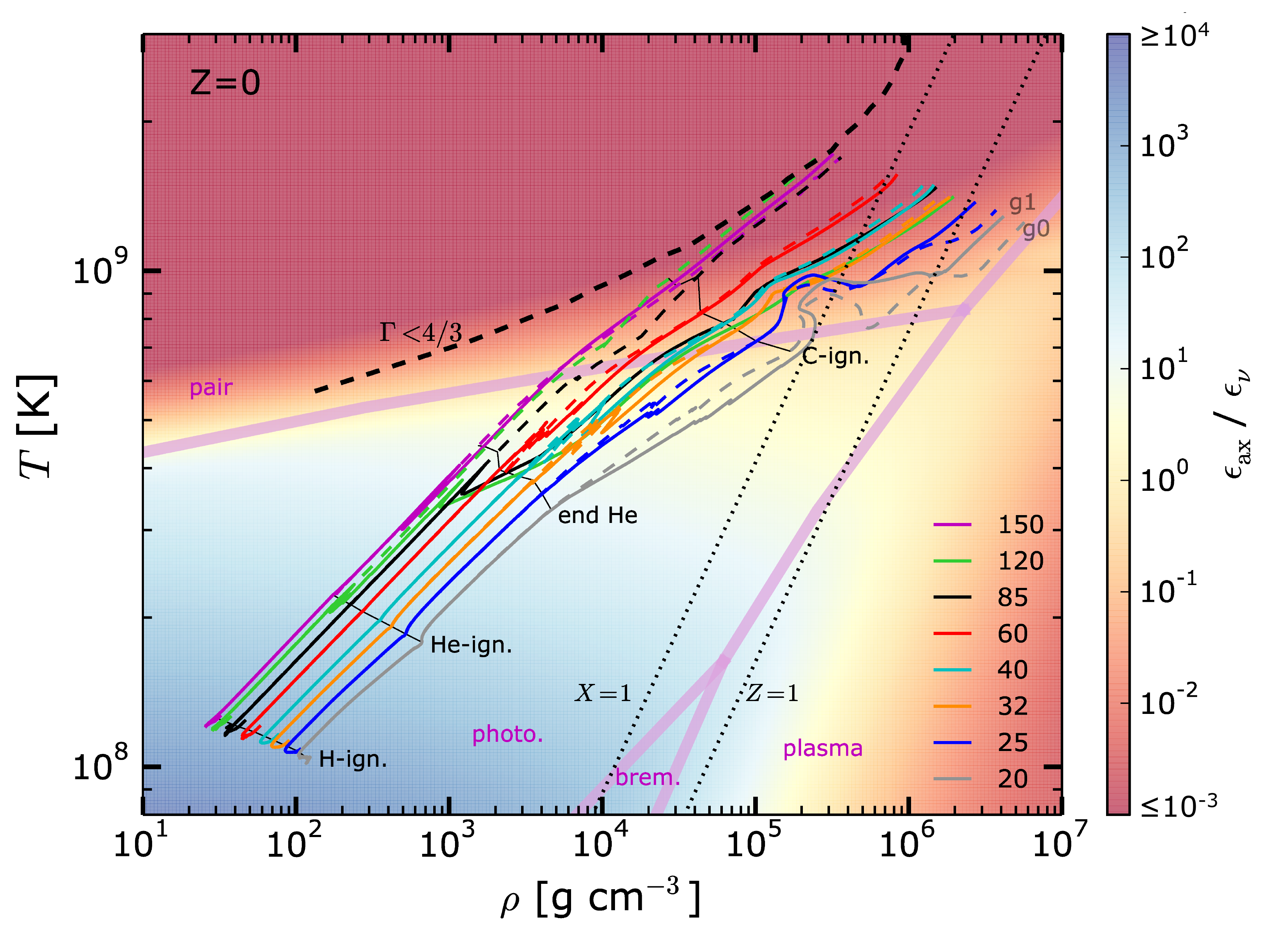} 
\caption{The colored lines shows the tracks for the computed Pop. III models in the central temperature vs. central density plane. Dashed lines show models without axion losses while solid lines show models with axion losses. The ignition of the different burning stages and the end of central helium burning are given for the models without axion losses. 
The heavy dashed $\Gamma < 4/3$ line indicates the zone of the diagram where electron-positron pair creations lower the adiabatic index below 4/3. The $X=1$ and $Z=1$ lines show the limit between the perfect gas and the completely-degenerate non-relativistic gas, for a pure hydrogen and pure metals mixtures. Light purple lines delimitate the different regions where a given neutrino source (pair, photoneutrino, bremsstrahlung and plasma) is dominating. The color map shows the ratio $\epsilon_{\rm ax}/\epsilon_{\rm \nu}$ where $\epsilon_{\rm \nu}$ is the sum of the four kind of neutrino losses mentioned above, computed according to \cite{Itoh89,Itoh96}. }
\label{f:trhopop3}
\end{center}
\end{figure*}

With the same physical ingredients used for solar metallicity models in the previous section, we have calculated Pop. III models of 20, 25, 32, 40, 60, 85, 120 and 150 $M_{\odot}$, with and without axions losses. The evolution is stopped at the end of core C-burning. We used the mass loss recipe of \cite{Vink01}. $\dot{M}$ depends on ($Z/Z_{\odot} )^{0.85}$ so that $\dot{M}$ should be 0 for our Pop. III models. However, we used $Z=10^{-4}$~$Z_{\odot}$ as a threshold value for the mass loss rate. This was done previously in \cite{Marigo03} and \cite{Ekstrom08}. Note that our models are computed at $Z=0$ strictly but we adopted for these models the mass loss rates given by a metallicity $Z=10^{-4}$ $Z_{\odot}$.
In some models, the opacity peak produced by the variation in the ionisation level of hydrogen decreases the Eddington luminosity $L_{\rm Edd} = 4\pi c GM / \kappa$ below the actual luminosity of the star. As a consequence, the external layers of such models exceed the Eddington luminosity. This unstable phase is not solvable with our hydrostatic approach. It is accounted for by increasing the mass-loss rate by a factor of 3, whenever the luminosity of any of the layers of the stellar envelope is higher that $5L_{\rm Edd}$ \citep{Ekstrom12}.

\subsection{Evolutionary tracks in the HR diagram and lifetimes}

The tracks in the HR diagram (Fig. \ref{f:htpop3}) are little affected by axion losses. The main differences between the two families of tracks arise close to the end of the evolution, when axion losses are important compared to other sources of energy loss. We see that in general, models with axions remain bluer in the HR diagram than the models without axions. Since axions add a new channel for evacuating the energy from the central regions, it removes energy that otherwise might be used to inflate the envelope and push the track in the HR diagram redward. In the previous section, we saw that axions could suppress a blue loop, thus causing a contraction of the envelope. Here we see that axions may reduce the expansion of the envelope. This difference arises here because we are considering the impact of axions at a different stage of the evolution. In the previous section, we were discussing the case of core He-burning star at the red supergiant stage. Here we are considering stars after the core He-burning phase still crossing the HR gap. In both cases axions remove energy. In the first case, it decreases the ability of the core to extend and hence the envelope to contract. In the second case, it removes energy released by contraction of the core that otherwise is used to expand the envelope. This is what happens for initial masses below $\sim$85 M$_\odot$. The 120g1 model loses $\sim$ 20 $M_{\odot}$ at the end of the core helium burning phase. Before mass ejection, $\log T_{\rm eff} \sim 3.8$ (solid green line on Fig.~\ref{f:htpop3}). Then, while losing mass, $\log T_{\rm eff}$ decreases until reaching a value of about 4.1. This model loses more mass than the 120g0 model (120 $M_{\odot}$ without axion) because it stays in a domain of the HR diagram where the luminosity becomes supra-Eddington for a longer time in some outer layers of the star.

The differences between the tracks in the HR diagram remain very modest. We computed different 85g1 models using different timesteps near the end of the evolution. This leads to slightly different final Teff's. The scatter is of the order of the difference in $T_{\rm eff}$ between the 85g0 and 85g1 models shown in Fig.\ref{f:htpop3}. In any case, such differences could not be used to constrain the presence of axions. This is true for at least two reasons. (1) Pop. III stars cannot be directly observed and thus they cannot be placed into a HR diagram.  With the James Webb Telescope, it will be possible on the other hand to detect supernovae from Pop. III stars. From the evolution of their early light curves (during the rise time), it might be possible to obtain some indications on the radius of the core collapse supernova progenitors~\citep[see e.g.][]{Nakar10,Bersten12, Dessart13, Morozova16}. However, this radius does not depend only on the presence of the axions. It also depends on convection, mixing in radiative layers, opacities, mass losses and thus there is little hope at the moment to use such a channel for constraining the physics of axions. (2) Axions may change somewhat the lifetimes of stars. Indirectly this may have an impact on the ionizing power of Pop. III stars. Changing the MS lifetime for instance, will change the duration of the phase when UV photos are emitted by the star. The impact of axions on lifetimes is shown in Fig. \ref{f:mtau}.  We see that axions shorten the Main-Sequence lifetime by less than 3\% and reduce the core helium burning lifetimes by $7-10$ \%. As already mentioned above, this simply reflects the fact that the effects of axions are the most marked when temperature increases, {\it i.e.} in the more advanced phases of the evolution. Again, here the effects are very modest and of the same order of magnitude as changes due to other uncertain physical ingredients of the models. A consequence of a shorter helium lifetime is that the $^{12}$C($\alpha$,$\gamma$)$^{16}$O reaction has less time to operate during core helium burning so that the $^{12}$C/$^{16}$O ratio at the end of central helium burning is slightly higher ($\lesssim 5$ \%) for models with shorter helium lifetimes, hence with axion losses.

\begin{figure}[t]
\begin{center}
 \includegraphics[width=.485\textwidth]{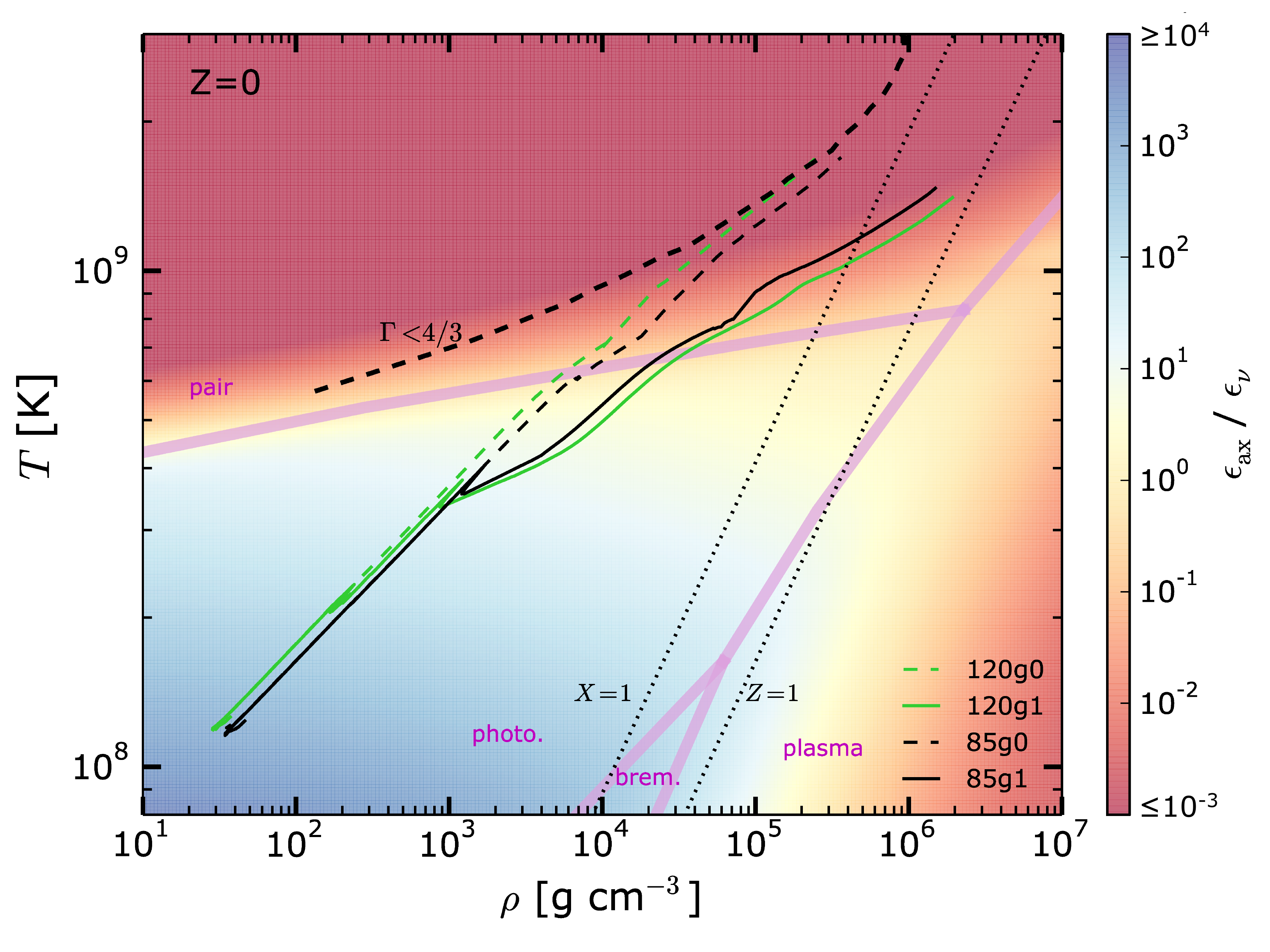} 
\caption{Same as Fig. \ref{f:trhopop3} but with only the 85 and 120 M$_{\odot}$ models.}
\label{f:trhopop3_2}
\end{center}
\end{figure}

\subsection{Central conditions}\label{sec42}

The colored lines of Figs. \ref{f:trhopop3} and \ref{f:trhopop3_2} show the evolution in the ($\log T_c$, $\log \rho_c$) diagram where $T_c$ and $\rho_c$ are respectively the central temperature and density. The color map shows the ratio $\epsilon_{\rm ax}/\epsilon_{\nu}$, between axion and neutrino energy losses. 
It shows that axions remove more energy than neutrinos from the central regions before the beginning of the core C-burning phase. After C-ignition in the center, neutrinos dominate. Thus, we can expect that axions will have their most important effects before the core C-burning phase. 
However, for axions to have a strong impact on the models, it is not sufficient that energy losses by axions are more important than neutrinos, energy losses due to axions must also be important with respect to the energy released either by nuclear reactions or gravitational contraction.

\begin{figure}[t]
   \centering
      \includegraphics[width=.485\textwidth]{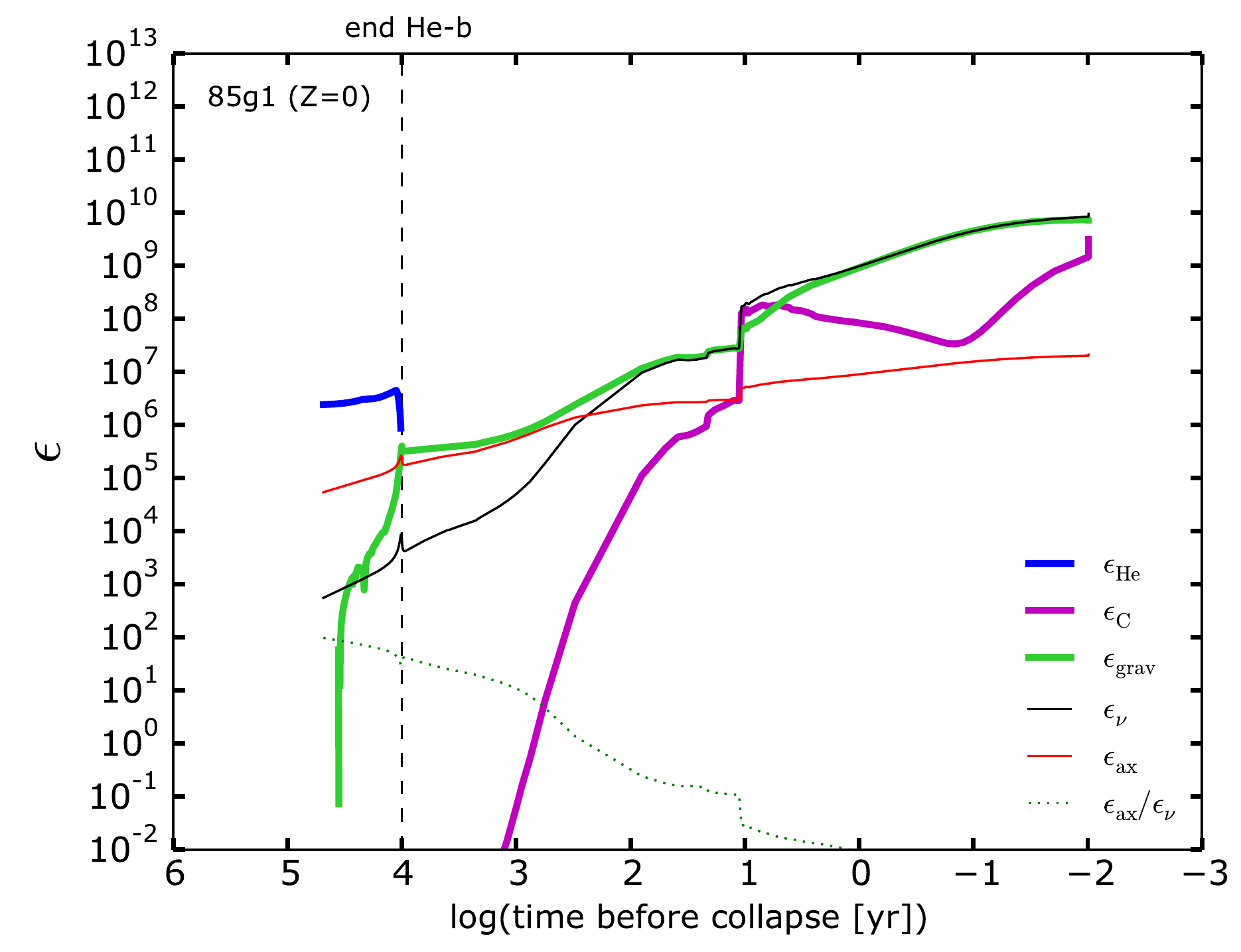}
   \caption{Energy generated by helium burning ($\epsilon_{\rm He}$), carbon burning ($\epsilon_{\rm C}$), gravitational energy ($\epsilon_{\rm grav}$), energy of the neutrinos ($\epsilon_{\nu}$) and of the axions ($\epsilon_{\rm ax}$). Thick lines represent sources of energy, thin lines represent sinks of energy. Dotted line shows the ratio $\epsilon_{\rm ax}$ / $\epsilon_{\rm \nu}$. The vertical dashed line denotes the end of the core helium burning phase.}
\label{f:teps}
    \end{figure}

During the core H-burning phase, the energy is mainly transported by radiation and convection. Axion cooling has little effect. The same is also true, although to a smaller extent, during the core He-burning phase. The only phase where strong differences may appear due to axions is during the transition between the end of the core He-burning phase and the beginning of the core C-burning phase. During that phase, energy in the central part is produced by core contraction (see the green line in Fig.~\ref{f:teps}). After the end of the core He-burning phase, the energy released by the contraction of the core is taken away by axions (between the abscissa 4 and 2.5 in Fig.~\ref{f:teps}). This is the phase during which axions may induce some significant changes in the models.

   \begin{figure}[t]
   \centering
      \includegraphics[width=.485\textwidth]{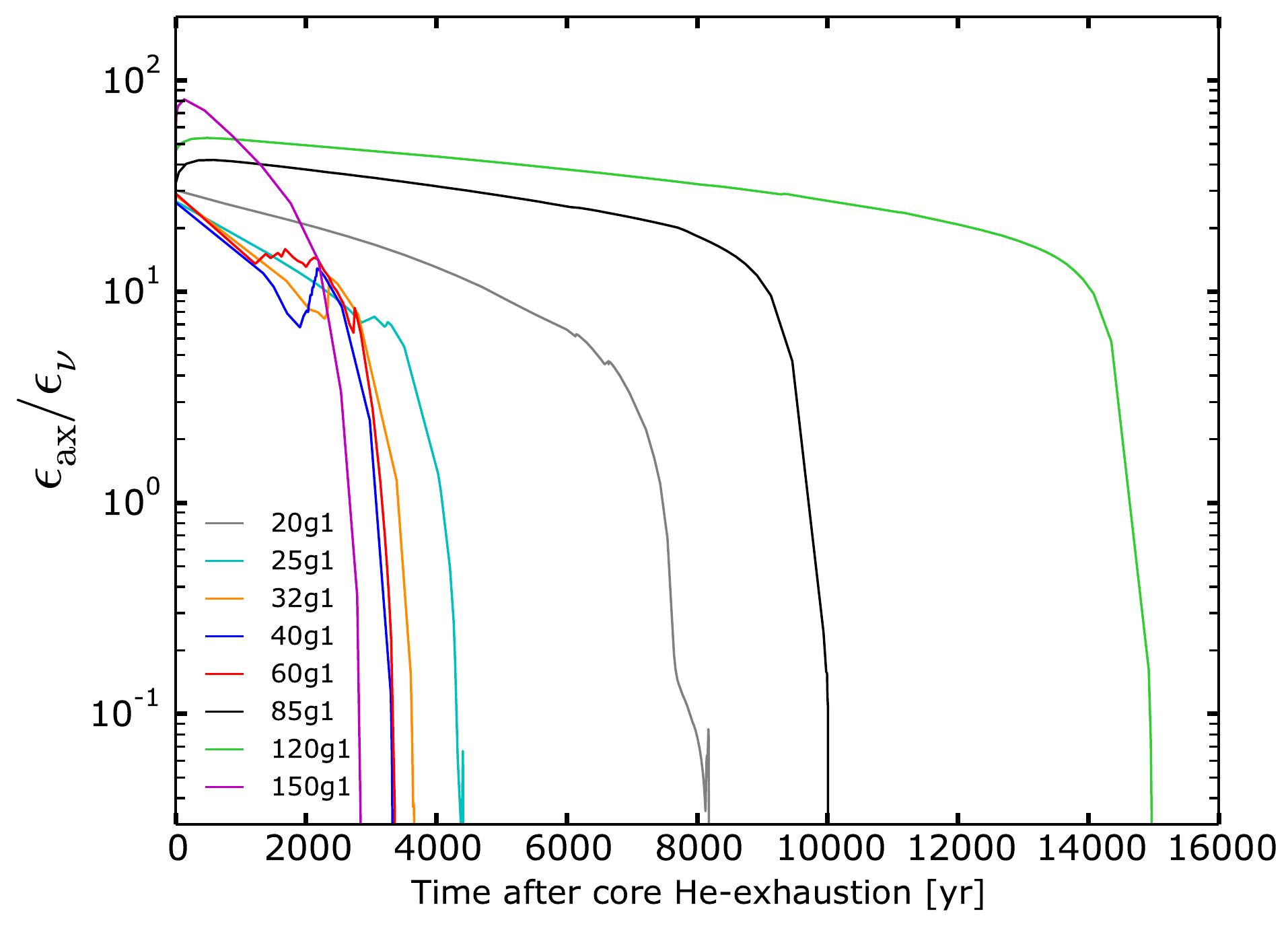}
   \caption{$\epsilon_{\rm ax} / \epsilon_{\rm \nu}$ ratio in the core of Pop. III models as a function of the time after the end of central helium burning.}
\label{f:eaxenu}
    \end{figure}

For most of the models, axions do not have a large effect. There is an exception however in the cases of the 85 and 120 M$_\odot$ models.
As can be better seen in Fig.~\ref{f:trhopop3_2}, the central density and temperature of the 85g1 (solid black line) and 120g1 (solid green line) models significantly deviates from the models without axions. After the core He-burning phase, the tracks with axions join respectively the tracks of the 40 (cyan tracks) and 32 (yellow tracks) $M_{\odot}$ models (see Fig.~\ref{f:trhopop3}), and follow these tracks until the end of their evolution. In other words, they follow the evolution of lower initial mass stars.

In these models, larger amount of energy can be evacuated from the central regions than in models with only neutrinos. This decreases the content of entropy in the core and thus will make the central region more sensitive to degeneracy effects. This explains why the tracks with axions approaches in a more rapid way the line separating the non-degenerate region from the degenerate one (see the lines
labeled $X$, $Z$=1 in Fig.~\ref{f:trhopop3_2}).

   \begin{figure*}[t]
   \centering
   \begin{minipage}[c]{.49\linewidth}
       \includegraphics[scale=0.25]{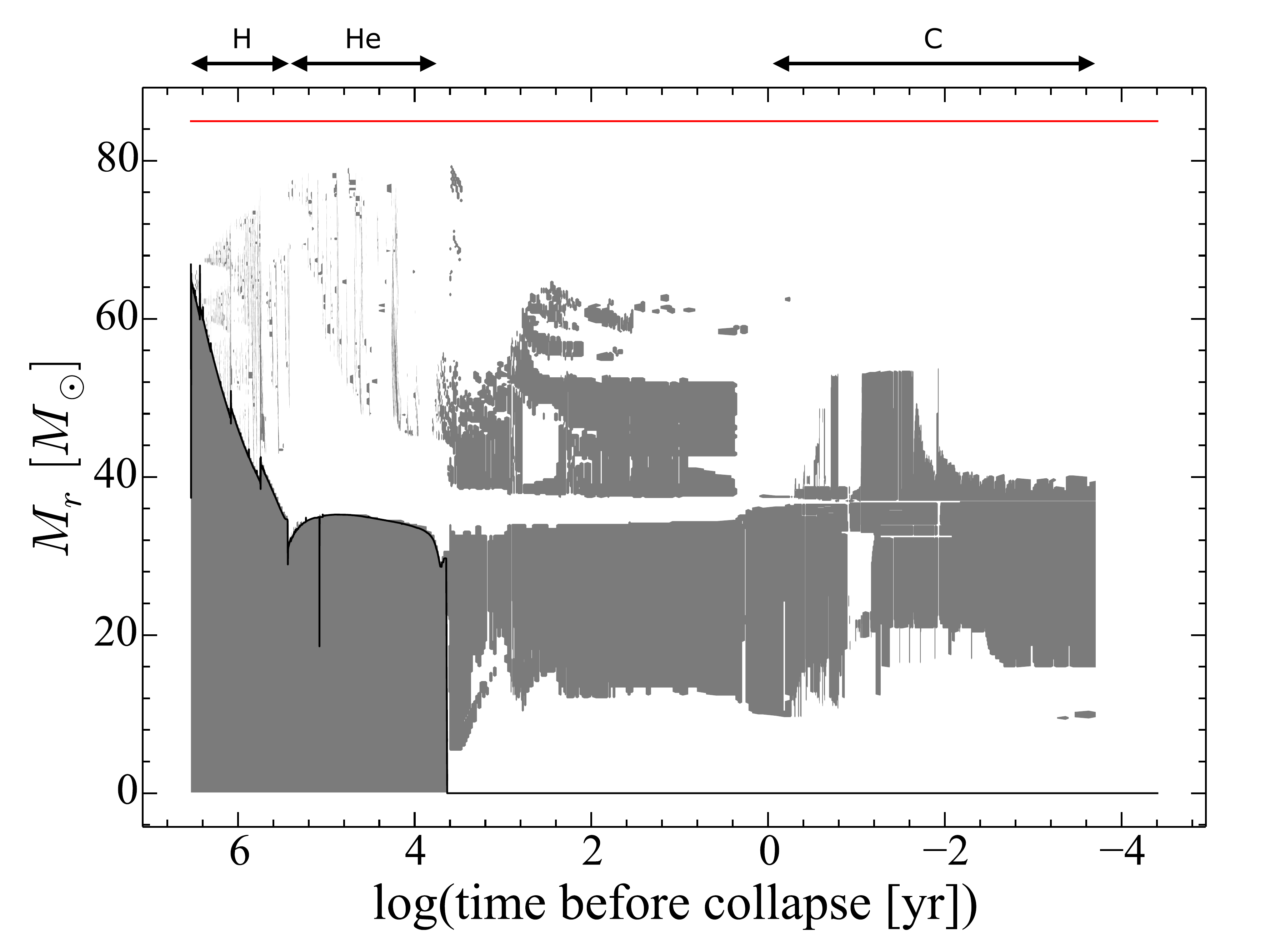}
   \end{minipage}
   \begin{minipage}[c]{.49\linewidth}
       \includegraphics[scale=0.25]{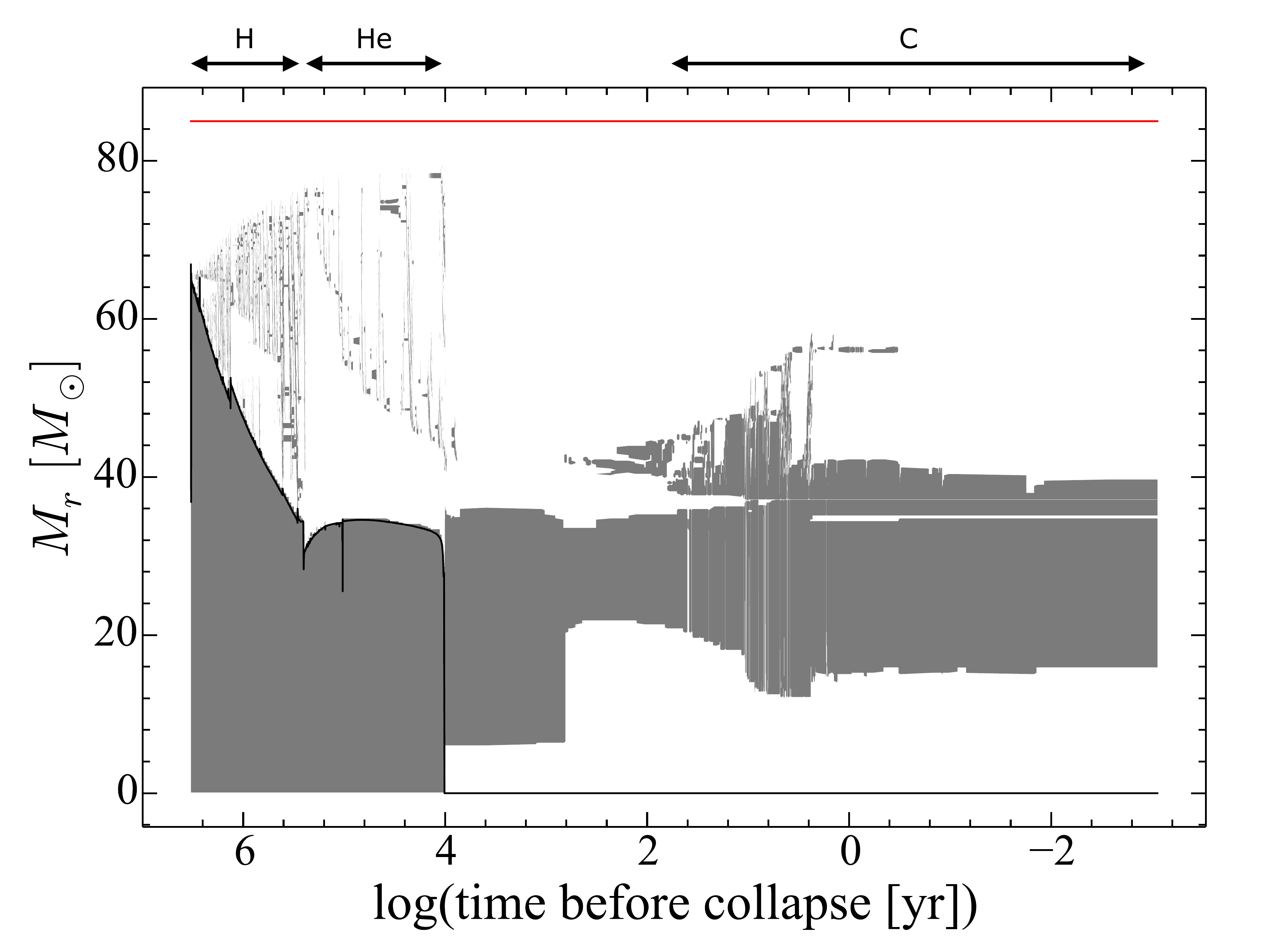}
   \end{minipage}
   \caption{Kippenhahn diagram of the 85g0 (left) and 85g1 (right) models. Grey areas represent the convective zones. The red line on the top shows the total mass. The core hydrogen, helium and carbon burning phases are indicated at the top of the panels.}
\label{f:kip}
    \end{figure*}

Surprisingly, this behavior disappears above and below the range 85-120 M$_\odot$. Outside this specific mass range, axions no longer have important effects and the tracks are only slightly changed when axions are considered. To understand why we have such a situation we note the following:
\begin{itemize}
\item As indicated above, axions dominates the process of energy removal from the central regions just after the core He-burning phase, until neutrinos become dominant. The durations of that phase, obtained in the different initial mass models computed with axions, is shown in Fig.~\ref{f:eaxenu}. We see that the longest axion-dominated phases occur for the 85 and 120 M$_\odot$ models. Thus this is consistent with the fact that axions have the largest impact in these models. The 20 M$_\odot$ model also has a rather long axion-dominated phase. A careful examination of Fig.~\ref{f:trhopop3} indicates that the track is more shifted than the other towards lower densities at a given temperature. This shift however disappears after C-ignition in the core.
\item The duration of the axion-dominated phase can be estimated to first order as $G M^2_{\rm core}/(R_{\rm core} L_{\rm axion})$, where $M_{\rm core}$, $R_{\rm core}$ are respectively the mass and the radius of the core at the end of the core He-burning phase and $L_{\rm axion}$ is the axion luminosity. This equation just expresses the fact that during that phase the energy source is mainly the gravitational energy, {\it i.e.} $G M^2_{\rm core}/R_{\rm core}$. The gravitational energy increases with the mass. The axion luminosity increases rapidly with the temperature and the temperature also increases with the mass. Therefore both the numerator and the denominator increase with the mass. This makes an easy prediction of how the ratio will vary as a function of mass difficult. Present numerical models tell us that for most of the initial masses,  this ratio keeps a nearly constant value. For a few initial masses, this ratio is however larger.
\end{itemize}

\subsection{Internal structure of the 85 and 120 $M_{\odot}$ models}

The structure evolution of the of the 85g1 and 120g1 models show striking differences compared to the 85g0 and 120g0 models.
This is illustrated in Fig.~\ref{f:kip} which compares the evolution as a function of time of the convective regions (grey areas) in the 85g0 and 85g1 models. We see that the core hydrogen and helium burning phases are very similar in both models. Differences appear after the core He-burning phase. In the model without axions, we have two  convective burning shells during the whole post He-burning phase. The outer one is the convective H-shell burning and the inner one the convective He-shell burning. In the models with axions, before C-ignition, there is only one convective burning shell, the He-burning one. We see also that during that phase the convective He-burning shell in the model with axions is more extended than in the model without axions. 
This is because in models with axions losses, more energy is removed from the He-shell and below. As a consequence, the star contracts more in that region. This tends to make this region warmer. This boosts the He-burning energy generation in the region where helium is still present and hence makes the convective shell associated with that burning more extended.

Because of the additional energy removed by axions, the cores of the 85g1 and 120g1 models contract more compared to the cores of the 85g0 and 120g0 models. This leads to a high central density that makes the cores of the 120g1 and 85g1 models partially degenerate.
In that case the energy released by contraction is in part used to push some electrons to occupy higher energy quantum levels. This energy is not used to increase the thermal energy. This implies that the 85g1 and 120g1 models end their evolution with a lower central temperature compared to the 85g0 and 120g0 models (see Fig. \ref{f:trhopop3_2}). We recall that on the other hand, the 85g1 and 120g1 models end their evolution with a higher temperature in the He-shell compared to the 85g0 and 120g0 models (c.f. discussion in the present Section).

The structure of such models, during that stage (in between helium and carbon burning), are nevertheless sensitive to other parameters, like the resolution (number of shells). We have computed again the 85g1 model with a different resolution and the convective pattern from abscissa $\sim 4$ in Fig. \ref{f:kip} (right panel) is different. This illustrates once again that other parameters can impact the structure of the models, during specific phases. What remains, nevertheless, is that the effect of the axions is the strongest during that stage (c.f. Fig. \ref{f:eaxenu} and discussion). This is during that stage that the axions are the most likely to have an impact on the evolution/structure of the star.

Also, the structure changes mentioned above are present only for a period of time. They appear as a transitory reaction of the model to the increased loss of energy in the central regions. In the end, the models converge towards structures that are again very similar (compare the two structures shown in Fig.~\ref{f:kip} at the very end of the evolution). The final masses of the helium and carbon-oxygen cores are little affected by the effect of axions (see table~\ref{table:1}).

\begin{table}
\scriptsize{
\caption{Masses of the helium and carbon-oxygen cores at the end of the core C-burning phase. \label{table:1}} 
\begin{center}
\resizebox{8.5cm}{!} {
\begin{tabular}{ll|cccccccc} 
\hline
\hline
 $M_{\rm ini}$ & [$M_{\odot}$] &  20  & 25 & 32 & 40 & 60 & 85 & 120 & 150 \\
\hline
 & & \multicolumn{8}{c}{\textbf{$g_{\rm 10} = 0$}}\\
 $M_{\rm \alpha}$ & [$M_{\odot}$] &  3.93 & 5.37 & 8.45 & 12.08 & 21.94 & 36.94 & 54.82 & 67.81 \\
 $M_{\rm CO}$ & [$M_{\odot}$] &  3.42  & 5.36 & 8.35 & 12.08 & 21.90 & 36.86 & 54.82 & 67.70 \\
\hline
 & & \multicolumn{8}{c}{\textbf{$g_{\rm 10} = 1$}}\\
 $M_{\rm \alpha}$ & [$M_{\odot}$] &  3.35 & 5.47 & 8.25 & 12.07 & 22.14 & 37.06 & 51.53 & 67.95 \\
 $M_{\rm CO}$ & [$M_{\odot}$] &  3.35  & 5.42 & 8.24 & 11.98 & 22.14 & 37.02 & 50.15 & 67.92 \\
\hline
\end{tabular}
}
\end{center}
}
\end{table}

It is interesting to note that although the 85 and 120 M$_\odot$ stars with axions would have very similar masses for the He and CO cores compared to their siblings without axions, they nevertheless show central conditions that are very different. As seen in Fig.~\ref{f:trhopop3_2}, they show much higher densities at a given temperature, thus, as already underlined above, the models with axions are much more sensitive to degeneracy effects. Whether or not this produce some flashes or even an explosion in the presupernova phases remain to be explored.


Recently, \cite{Woosley17} has investigated the final evolution of $70-140$ $M_{\odot}$ stars. He finds that such stars should experience Pulsational Pair Instability Supernovae (PPISN). PPISN would occur if the final helium core is more massive than 30 $M_{\odot}$. For helium cores above 62 $M_{\odot}$, the star is disrupted as a single pulse (Pair Instability Supernovae). Our 85, 120 and 150 $M_{\odot}$ models have helium cores between $\sim 37$ and 68 $M_{\odot}$ (see Table \ref{table:1}). They might experience PPISN (or PISN for helium cores above 62 $M_{\odot}$). However, the axion cooling might prevent our 85 and 120 $M_{\odot}$ models from entering in the unstable regime (see Fig. \ref{f:trhopop3} and \ref{f:trhopop3_2}). If so, such stars could produce black holes in the mass range where no black hole is usually expected, between 64 and 133 $M_{\odot}$ \citep{Heger02}. This could induce a potential signature in gravitationnal waves. Although still far from being able to use stellar models for testing the existence of axions, this possible axion signature has to be kept in mind for the future.

\subsection{Impact of the axion cooling on nucleosynthesis}

Pop. III stars lose very little mass through stellar winds \citep[although fast rotation may change this picture, see][]{Ekstrom08, Ekstrom08proc}. Thus the only way they can contribute to nucleosynthesis
would be through supernova ejecta. As discussed above, the two masses for which the impact of axions are the strongest are the 85 and 120 M$_\odot$. If these two stars would end their life producing a black-hole with very little or no mass ejecta, then their contribution in enriching the interstellar medium would likely be null. However, a successful explosion is also a possible scenario \citep[e.g.][]{Ertl16, Sukhbold16}.
The discussion below investigates what could be the impact of axions if some material is ejected.

\begin{figure}[t]
\begin{center}
 \includegraphics[width=.485\textwidth]{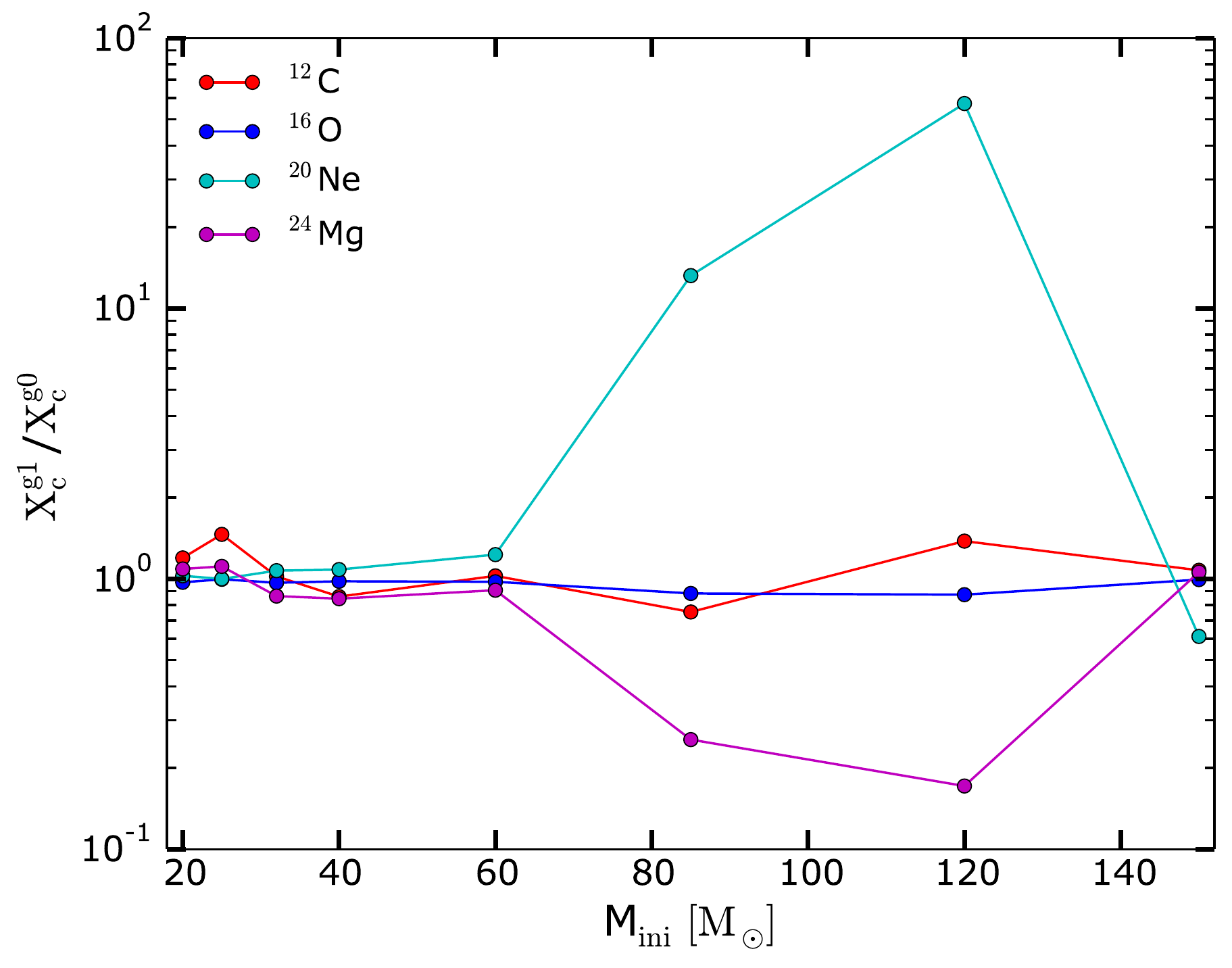} 
\caption{Ratio of central abundances at the end of core C-burning between models with axions losses ($\rm X^{g1}_{c}$) and models without axion losses ($\rm X^{g0}_{c}$).}
\label{f:mxcg}
\end{center}
\end{figure}

Figure~\ref{f:mxcg} shows central abundance ratios at the end of core C-burning between models with and without axions losses. The most abundant species are shown. The central abundances of $20-60$ M$_{\odot}$ models are little affected by axion losses. Over the whole range of masses considered, $^{12}$C and $^{16}$O are little affected. $^{20}$Ne and $^{24}$Mg are affected in the range $80-140$ M$_{\odot}$. The central $^{20}$Ne ($^{24}$Mg) abundance is $\sim 60$ times higher ($\sim 5$ times lower) in the 120g1 model compared to the 120g0 model.
This is due to the higher central temperature in the 85g0 (120g0) compared to the 85g1 (120g1) model (c.f. discussion in Sect.~\ref{sec42}). 
In the case of the 85g0 and 120g0 models, the central temperature reaches values $> 1.4$ GK, where the $^{20}$Ne($\alpha$,$\gamma$)$^{24}$Mg channel becomes stronger than the $^{16}$O($\alpha$,$\gamma$)$^{20}$Ne channel so that significantly more $^{24}$Mg is synthesized to the detriment of $^{20}$Ne. 
However, the central abundances are not sufficient for determining the yield of an element. The complete chemical structure should be inspected.

   \begin{figure*}
   \centering
   \begin{minipage}[c]{.49\linewidth}
       \includegraphics[width=1.08\textwidth]{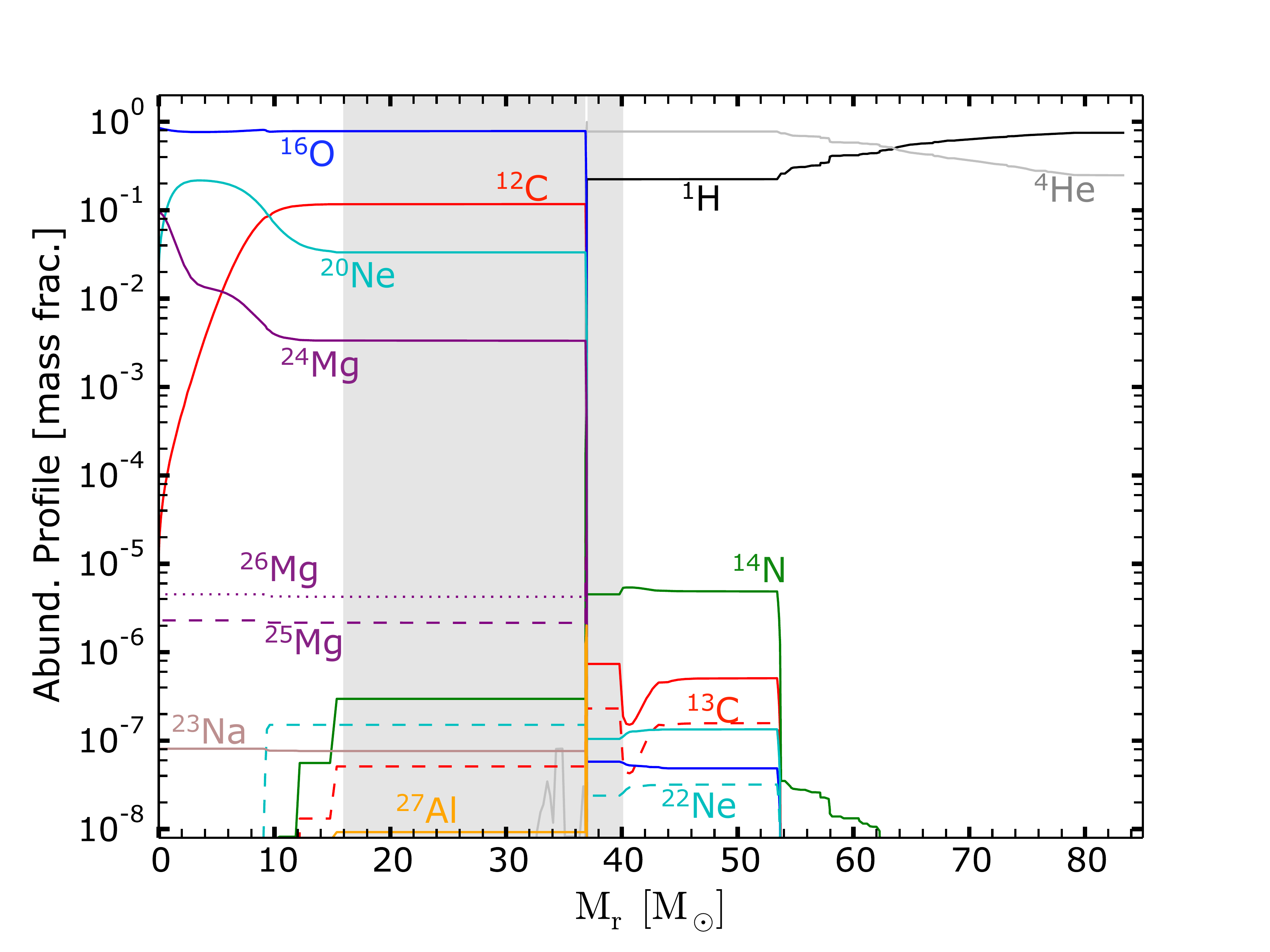}
   \end{minipage}
   \begin{minipage}[c]{.49\linewidth}
       \includegraphics[width=1.08\textwidth]{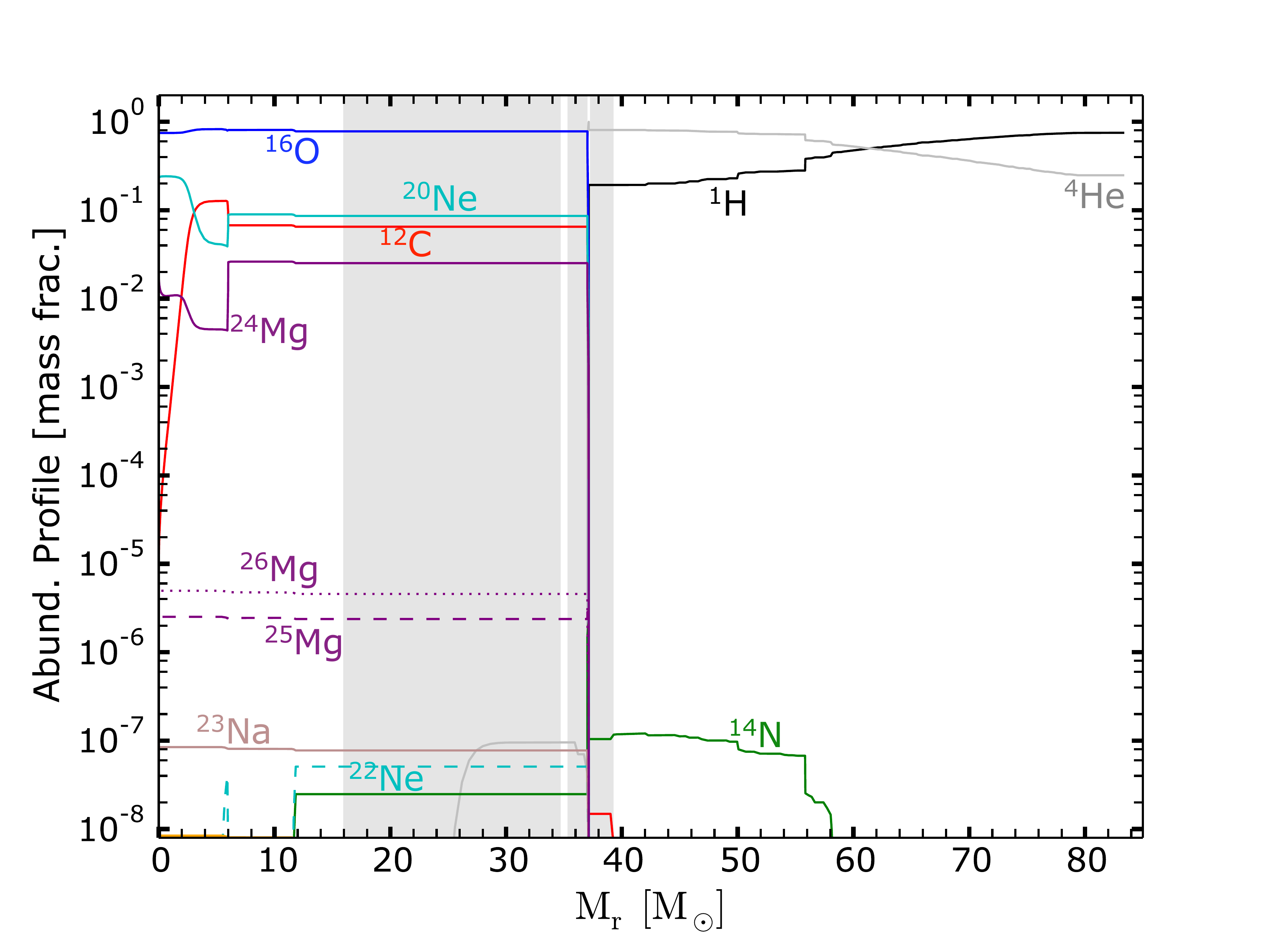}
   \end{minipage}
   \caption{Abundance profiles of the 85g0 (left) and 85g1 (right) models at the end of core carbon burning (central mass fraction of $^{12}$C below $10^{-5}$). The shaded area show the convective zones. }
\label{f:abund}
    \end{figure*}

Figure~\ref{f:abund} shows the complete chemical structure obtained in the two 85 M$_\odot$ models with and without axions. These structures are also representative of the 120 M$_\odot$ models. For the other models, the axions do not lead to significant differences in the final abundance profiles and we will not discuss them further. We see in Fig.~\ref{f:abund} that the distribution of oxygen is very similar in both models and thus we do not expect important effects on the stellar yield of that element. For $^{12}$C, we see some differences: the zone where carbon is depleted is smaller for the 85g1 model. This is because of the more efficient cooling due to axion emissions, prior the core carbon-burning phase. This implies that a smaller fraction of the core reaches the conditions needed for carbon burning to occur. The $^{14}$N profiles are a bit different but the mass fraction stays very small, below $10^{-5}$ in any case. This difference is not significant.

In the He-burning shell, the abundance of $^{20}$Ne and $^{24}$Mg are larger in the model with axions. These differences reflect the higher temperatures reached in the He-burning shell in the model with axions (c.f. discussion in Sect.~\ref{sec42}). 
On the whole, we see that in case these layers associated to the He-shell burning are ejected, this may boost the yield by a factor ten with respect to the yield of these two isotopes obtained in the corresponding models without axions. However, this will occur in a relatively short mass range and thus the global effect on the yields integrated over an initial mass function where these massive stars are far out-numbered
by lower mass stars, will remain quite modest.

\section{Conclusions}\label{sec5}

We have explored the impact of axions in massive stars at solar metallicity and in Pop. III stars. 
In solar metallicity stars, we have confirmed that the axion coupling to photons might be constrained
due to the disappearance of the blue-loop phase in stars with masses between 8 and 12 M$_\odot$. However, caution is required because blue loops are very sensitive to many other uncertain physical processes (e.g. core convection and its efficiency and boundary mixing), which can also remove the blue loops from a stellar evolutionary track without need for axion cooling.
In Pop. III stars, we have shown that the effects of axions on stellar evolution are very modest and are hardly observable, mainly because uncertainties in other parameters of stellar models produce at least as large effects. 

The most spectacular effect of axions explored in the present work is on the physical conditions obtained at the end of the evolution for Pop. III stars with masses between about 80 and 130 M$_\odot$. These stars present cores with some level of degeneracy and the final fate such objects is not clear. \cite{Sukhbold16} have investigated the final fate of $9 - 120$ M$_\odot$ solar metallicity models. Above 30 M$_{\odot}$, most of the models do not explode and form black holes. The effect of axions might change this picture.
This clearly needs further investigation especially as we approach the era of observations with the James Webb Telescope that might detect supernovae in the entire observable Universe. Such stars might give rise to a type of explosive event with a clear non-ambiguous signature inherited from the particular structure of their progenitors.

On a broader perspective, this work shows that we are still far from a situation where Pop. III stellar models can be used as physics laboratory to test for the presence of axions and even further from the situation where stars might constrain their properties. This shows the importance of continuing our efforts in stellar physics, making stellar models still more precise and reliable  in order that real stars can be used as physics laboratories opening new windows on questions at the frontier of physics, in domain of temperatures and densities that are not reachable in a laboratory.

\begin{acknowledgements} 
We would like to thank Jordi Isern for useful discussions during the early stage of this project. The work K. A. O. was supported in part by DOE grant DE-SC0011842 at the University of Minnesota.
The work of E. V. has been carried out at the ILP LABEX (under reference ANR-10-LABX-63) supported by French state funds managed by the ANR within the Investissements d'Avenir programme under reference ANR-11-IDEX-0004-02. 
\end{acknowledgements}

\bibliographystyle{aa}
\bibliography{biblio}

\begin{thebibliography}{55}
\expandafter\ifx\csname natexlab\endcsname\relax\def\natexlab#1{#1}\fi

\bibitem[{{Abbott} \& {Sikivie}(1983)}]{Abbott83}
{Abbott}, L.~F. \& {Sikivie}, P. 1983, Physics Letters B, 120, 133

\bibitem[{{Aoyama} \& {Suzuki}(2015)}]{Aoyama15}
{Aoyama}, S. \& {Suzuki}, T.~K. 2015, \prd, 92, 063016

\bibitem[{{Asplund} {et~al.}(2005){Asplund}, {Grevesse}, \&
  {Sauval}}]{Asplund05}
{Asplund}, M., {Grevesse}, N., \& {Sauval}, A.~J. 2005, in Astronomical Society
  of the Pacific Conference Series, Vol. 336, Cosmic Abundances as Records of
  Stellar Evolution and Nucleosynthesis, ed. T.~G. {Barnes}, III \& F.~N.
  {Bash}, 25

\bibitem[{{Ayala} {et~al.}(2014){Ayala}, {Dom{\'{\i}}nguez}, {Giannotti},
  {Mirizzi}, \& {Straniero}}]{Ayala14}
{Ayala}, A., {Dom{\'{\i}}nguez}, I., {Giannotti}, M., {Mirizzi}, A., \&
  {Straniero}, O. 2014, Physical Review Letters, 113, 191302

\bibitem[{{Bagnaschi} {et~al.}(2015){Bagnaschi}, {Buchmueller}, {Cavanaugh},
  {Citron}, {De Roeck}, {Dolan}, {Ellis}, {Fl{\"a}cher}, {Heinemeyer},
  {Isidori}, {Malik}, {Mart{\'{\i}}nez Santos}, {Olive}, {Sakurai}, {de Vries},
  \& {Weiglein}}]{Bagnaschi15}
{Bagnaschi}, E.~A., {Buchmueller}, O., {Cavanaugh}, R., {et~al.} 2015, European
  Physical Journal C, 75, 500

\bibitem[{{Bersten} {et~al.}(2012){Bersten}, {Benvenuto}, {Nomoto}, {Ergon},
  {Folatelli}, {Sollerman}, {Benetti}, {Botticella}, {Fraser}, {Kotak},
  {Maeda}, {Ochner}, \& {Tomasella}}]{Bersten12}
{Bersten}, M.~C., {Benvenuto}, O.~G., {Nomoto}, K., {et~al.} 2012, \apj, 757,
  31

\bibitem[{{Bromm}(2013)}]{Bromm13}
{Bromm}, V. 2013, Reports on Progress in Physics, 76, 112901

\bibitem[{{Burrows} {et~al.}(1990){Burrows}, {Ressell}, \&
  {Turner}}]{Burrows90}
{Burrows}, A., {Ressell}, M.~T., \& {Turner}, M.~S. 1990, \prd, 42, 3297

\bibitem[{{Cunha} {et~al.}(2006){Cunha}, {Hubeny}, \& {Lanz}}]{Cunha06}
{Cunha}, K., {Hubeny}, I., \& {Lanz}, T. 2006, \apjl, 647, L143

\bibitem[{{Dearborn} {et~al.}(1986){Dearborn}, {Schramm}, \&
  {Steigman}}]{Dearborn86}
{Dearborn}, D.~S.~P., {Schramm}, D.~N., \& {Steigman}, G. 1986, Physical Review
  Letters, 56, 26

\bibitem[{{Dessart} {et~al.}(2013){Dessart}, {Hillier}, {Waldman}, \&
  {Livne}}]{Dessart13}
{Dessart}, L., {Hillier}, D.~J., {Waldman}, R., \& {Livne}, E. 2013, \mnras,
  433, 1745

\bibitem[{{Dine} \& {Fischler}(1983)}]{Dine83}
{Dine}, M. \& {Fischler}, W. 1983, Physics Letters B, 120, 137

\bibitem[{{Dine} {et~al.}(1981){Dine}, {Fischler}, \& {Srednicki}}]{Dine81}
{Dine}, M., {Fischler}, W., \& {Srednicki}, M. 1981, Physics Letters B, 104,
  199

\bibitem[{{Ekstr{\"o}m} {et~al.}(2012){Ekstr{\"o}m}, {Georgy}, {Eggenberger},
  {Meynet}, {Mowlavi}, {Wyttenbach}, {Granada}, {Decressin}, {Hirschi},
  {Frischknecht}, {Charbonnel}, \& {Maeder}}]{Ekstrom12}
{Ekstr{\"o}m}, S., {Georgy}, C., {Eggenberger}, P., {et~al.} 2012, \aap, 537,
  A146

\bibitem[{{Ekstr{\"o}m} {et~al.}(2008{\natexlab{a}}){Ekstr{\"o}m}, {Meynet},
  {Chiappini}, {Hirschi}, \& {Maeder}}]{Ekstrom08}
{Ekstr{\"o}m}, S., {Meynet}, G., {Chiappini}, C., {Hirschi}, R., \& {Maeder},
  A. 2008{\natexlab{a}}, \aap, 489, 685

\bibitem[{{Ekstr{\"o}m} {et~al.}(2008{\natexlab{b}}){Ekstr{\"o}m}, {Meynet}, \&
  {Maeder}}]{Ekstrom08proc}
{Ekstr{\"o}m}, S., {Meynet}, G., \& {Maeder}, A. 2008{\natexlab{b}}, in IAU
  Symposium, Vol. 250, Massive Stars as Cosmic Engines, ed. F.~{Bresolin},
  P.~A. {Crowther}, \& J.~{Puls}, 209--216

\bibitem[{{Ellis} {et~al.}(1984){Ellis}, {Hagelin}, {Nanopoulos}, {Olive}, \&
  {Srednicki}}]{Ellis84}
{Ellis}, J., {Hagelin}, J.~S., {Nanopoulos}, D.~V., {Olive}, K., \&
  {Srednicki}, M. 1984, Nuclear Physics B, 238, 453

\bibitem[{{Ellis} \& {Olive}(1987)}]{Ellis87}
{Ellis}, J. \& {Olive}, K.~A. 1987, Physics Letters B, 193, 525

\bibitem[{{Ertl} {et~al.}(2016){Ertl}, {Janka}, {Woosley}, {Sukhbold}, \&
  {Ugliano}}]{Ertl16}
{Ertl}, T., {Janka}, H.-T., {Woosley}, S.~E., {Sukhbold}, T., \& {Ugliano}, M.
  2016, \apj, 818, 124

\bibitem[{{Friedland} {et~al.}(2013){Friedland}, {Giannotti}, \&
  {Wise}}]{Friedland13}
{Friedland}, A., {Giannotti}, M., \& {Wise}, M. 2013, Physical Review Letters,
  110, 061101

\bibitem[{{Goldberg}(1983)}]{Goldberg83}
{Goldberg}, H. 1983, Physical Review Letters, 50, 1419

\bibitem[{{Heger} \& {Woosley}(2002)}]{Heger02}
{Heger}, A. \& {Woosley}, S.~E. 2002, \apj, 567, 532

\bibitem[{{Itoh} {et~al.}(1989){Itoh}, {Adachi}, {Nakagawa}, {Kohyama}, \&
  {Munakata}}]{Itoh89}
{Itoh}, N., {Adachi}, T., {Nakagawa}, M., {Kohyama}, Y., \& {Munakata}, H.
  1989, \apj, 339, 354

\bibitem[{{Itoh} {et~al.}(1996){Itoh}, {Hayashi}, {Nishikawa}, \&
  {Kohyama}}]{Itoh96}
{Itoh}, N., {Hayashi}, H., {Nishikawa}, A., \& {Kohyama}, Y. 1996, \apjs, 102,
  411

\bibitem[{{Kawasaki} \& {Nakayama}(2013)}]{Kawasaki13}
{Kawasaki}, M. \& {Nakayama}, K. 2013, Annual Review of Nuclear and Particle
  Science, 63, 69

\bibitem[{{Keil} {et~al.}(1997){Keil}, {Janka}, {Schramm}, {Sigl}, {Turner}, \&
  {Ellis}}]{Keil97}
{Keil}, W., {Janka}, H.-T., {Schramm}, D.~N., {et~al.} 1997, \prd, 56, 2419

\bibitem[{{Kim}(1979)}]{Kim79}
{Kim}, J.~E. 1979, Physical Review Letters, 43, 103

\bibitem[{{Kippenhahn} \& {Weigert}(1990)}]{Kippenhahn90}
{Kippenhahn}, R. \& {Weigert}, A. 1990, {Stellar Structure and Evolution}, 192

\bibitem[{{Lodders}(2003)}]{Lodders03}
{Lodders}, K. 2003, \apj, 591, 1220

\bibitem[{{Maeder} \& {Meynet}(2001)}]{Maeder01}
{Maeder}, A. \& {Meynet}, G. 2001, \aap, 373, 555

\bibitem[{{Marigo} {et~al.}(2003){Marigo}, {Chiosi}, \& {Kudritzki}}]{Marigo03}
{Marigo}, P., {Chiosi}, C., \& {Kudritzki}, R.-P. 2003, \aap, 399, 617

\bibitem[{{Marsh}(2016)}]{Marsh16}
{Marsh}, D.~J.~E. 2016, \physrep, 643, 1

\bibitem[{{Mayle} {et~al.}(1988){Mayle}, {Wilson}, {Ellis}, {Olive}, {Schramm},
  \& {Steigman}}]{Mayle88}
{Mayle}, R., {Wilson}, J.~R., {Ellis}, J., {et~al.} 1988, Physics Letters B,
  203, 188

\bibitem[{{Mayle} {et~al.}(1989){Mayle}, {Wilson}, {Ellis}, {Olive}, {Schramm},
  \& {Steigman}}]{Mayle89}
{Mayle}, R., {Wilson}, J.~R., {Ellis}, J., {et~al.} 1989, Physics Letters B,
  219, 515

\bibitem[{{Morozova} {et~al.}(2016){Morozova}, {Piro}, {Renzo}, \&
  {Ott}}]{Morozova16}
{Morozova}, V., {Piro}, A.~L., {Renzo}, M., \& {Ott}, C.~D. 2016, \apj, 829,
  109

\bibitem[{{Nakar} \& {Sari}(2010)}]{Nakar10}
{Nakar}, E. \& {Sari}, R. 2010, \apj, 725, 904

\bibitem[{{Peccei} \& {Quinn}(1977{\natexlab{a}})}]{Peccei77b}
{Peccei}, R.~D. \& {Quinn}, H.~R. 1977{\natexlab{a}}, \prd, 16, 1791

\bibitem[{{Peccei} \& {Quinn}(1977{\natexlab{b}})}]{Peccei77a}
{Peccei}, R.~D. \& {Quinn}, H.~R. 1977{\natexlab{b}}, Physical Review Letters,
  38, 1440

\bibitem[{{Preskill} {et~al.}(1983){Preskill}, {Wise}, \&
  {Wilczek}}]{Preskill83}
{Preskill}, J., {Wise}, M.~B., \& {Wilczek}, F. 1983, Physics Letters B, 120,
  127

\bibitem[{{Primakoff}(1951)}]{Primakoff51}
{Primakoff}, H. 1951, Physical Review, 81, 899

\bibitem[{{Raffelt} \& {Seckel}(1988)}]{Raffelt88}
{Raffelt}, G. \& {Seckel}, D. 1988, Physical Review Letters, 60, 1793

\bibitem[{{Raffelt} \& {Seckel}(1991)}]{Raffelt91}
{Raffelt}, G. \& {Seckel}, D. 1991, Physical Review Letters, 67, 2605

\bibitem[{{Raffelt} \& {Weiss}(1995)}]{Raffelt95}
{Raffelt}, G. \& {Weiss}, A. 1995, \prd, 51, 1495

\bibitem[{{Raffelt}(1986)}]{Raffelt86}
{Raffelt}, G.~G. 1986, \prd, 33, 897

\bibitem[{{Raffelt}(1990)}]{Raffelt90}
{Raffelt}, G.~G. 1990, \physrep, 198, 1

\bibitem[{{Raffelt}(2008)}]{Raffelt08}
{Raffelt}, G.~G. 2008, in Lecture Notes in Physics, Berlin Springer Verlag,
  Vol. 741, Axions, ed. M.~{Kuster}, G.~{Raffelt}, \& B.~{Beltr{\'a}n}, 51

\bibitem[{{Raffelt} \& {Dearborn}(1987)}]{Raffelt87}
{Raffelt}, G.~G. \& {Dearborn}, D.~S.~P. 1987, \prd, 36, 2211

\bibitem[{{Shifman} {et~al.}(1980){Shifman}, {Vainshtein}, \&
  {Zakharov}}]{Shifman80}
{Shifman}, M.~A., {Vainshtein}, A.~I., \& {Zakharov}, V.~I. 1980, Nuclear
  Physics B, 166, 493

\bibitem[{{Sukhbold} {et~al.}(2016){Sukhbold}, {Ertl}, {Woosley}, {Brown}, \&
  {Janka}}]{Sukhbold16}
{Sukhbold}, T., {Ertl}, T., {Woosley}, S.~E., {Brown}, J.~M., \& {Janka}, H.-T.
  2016, \apj, 821, 38

\bibitem[{{Vink} {et~al.}(2001){Vink}, {de Koter}, \& {Lamers}}]{Vink01}
{Vink}, J.~S., {de Koter}, A., \& {Lamers}, H.~J.~G.~L.~M. 2001, \aap, 369, 574

\bibitem[{{Walmswell} {et~al.}(2015){Walmswell}, {Tout}, \&
  {Eldridge}}]{Walmswell15}
{Walmswell}, J.~J., {Tout}, C.~A., \& {Eldridge}, J.~J. 2015, \mnras, 447, 2951

\bibitem[{{Weinberg}(1978)}]{Weinberg78}
{Weinberg}, S. 1978, Physical Review Letters, 40, 223

\bibitem[{{Wilczek}(1978)}]{Wilczek78}
{Wilczek}, F. 1978, Physical Review Letters, 40, 279

\bibitem[{{Woosley}(2017)}]{Woosley17}
{Woosley}, S.~E. 2017, \apj, 836, 244

\bibitem[{{Zhitnitskii}(1980)}]{Zhitnitskii80}
{Zhitnitskii}, A.~P. 1980, Sov. J. Nucl. Phys., 31, 260

\end{thebibliography}

\section*{Appendix A}

The energy loss rate is given by  Eq.~(3) of \cite{Friedland13} has the form
\begin{equation}
\epsilon_{\rm ax}=Kg_{10}^2T_8^7\rho_3^{-1}Z(\xi^2)
\label{q:lossx}
\end{equation}
where the function $Z(\xi^2)$ is identical to  $\xi^2f(\xi^2)$ in Eq.~\ref{q:loss}. 
However, the numerical value of the constant $K$, was found to be incorrect ($K\neq27.2$) in \cite{Friedland13}. Indeed, starting from Eq.~\ref{q:loss} and inserting the correct constants to obtain the proper dimensions, one obtains
\begin{equation}
\epsilon_{\rm ax}=\mathrm{c}\frac{g_{a\gamma}^2\left(\mathrm{k_B}T\right)^7}{\left(\hbar\mathrm{c}\right)^44\pi^2\rho}\xi^2f(\xi^2).
\label{q:epsb}
\end{equation}
For  $g_{a\gamma}=10^{-10}$~GeV$^{-1}$, $T=10^8$~K, $\rho=10^3$~g/cm$^3$, and using  $\mathrm{k_B}=8.617343\times10^{-14}$~GeV/K, $\hbar\mathrm{c}=197.326968\times10^{-16}$~GeV~cm,  $\mathrm{c}=299792458\times10^2$~cm/s and 1~GeV = 1.60217653$\times10^{-3}$~erg, one obtains the correct value,  namely $K$=283.16~erg/g/s. It is a factor of ten higher than in \cite{Aoyama15} and \cite{Friedland13} due to a propagated typo.  We confirmed that calculations by \cite{Aoyama15}  were done with the incorrect value and need to be re--calculated. Our value is also consistent with Fig. 8.6 in Raffelt's review~\citep{Raffelt90}. 

Nevertheless, since \cite{Friedland13} provide the energy loss (neutrinos plus axions) subroutine used in their MESA calculations  we analysed their modified MESA subroutine\footnote{\tt http://alexfriedland.com/papers/axion/} to check if they used the correct formula.  Combining Eqs.~\ref{q:xi} and \ref{q:debye}, one gets for $\xi^2$ ({\tt axioncsi}  in their code),
\begin{equation}
\xi^2=
\pi\alpha\left(\frac{\hbar\mathrm{c}}{\mathrm{k_B}T}\right)^3\mathcal{N}_A\rho\sum_i{Y_i}Z_i^2,
\label{q:xi2a}
\end{equation}
where $Y_i$ are the molar fractions $Y=X/A$ (mole/g). If  $T$ and $\rho$ are in K and g/cm$^3$, it leads to
\begin{equation}
\xi^2=
1.65769\times10^{20}\frac{\rho}{T^3}\sum_i{Y_i}Z_i^2,
\label{q:xi2b}
\end{equation}
with the numerical factor in agreement with the Friedland's code,
\begin{verbatim}
      axioncsi=  1.658d20*axionz2ye*Rho/T**3.
\end{verbatim}
Combining Eq.~\ref{q:epsb} and \ref{q:xi2a}, one obtains:
\begin{equation}
\epsilon_{\rm ax}=\alpha\frac{g_{a\gamma}^2\left(\mathrm{k_B}T\right)^4}{4\pi\hbar}
\mathcal{N}_Af(\xi^2)\sum_i{Y_i}Z_i^2.
\label{q:epsz}
\end{equation}
It is equivalent to Eq.~\ref{q:loss}, without the artificial $\rho$ dependence, now only present in $f(\xi^2)$. With $T$ in K, it gives
\begin{equation}
\epsilon_{\rm a}=4.694\times10^{-31}f(\xi^2)\;g_{10}^2T^4\sum_i{Y_i}Z_i^2 (\mathrm{erg/g/s})
\label{q:lossy}
\end{equation}
with, again a numerical factor in agreement with the Friedland's code,
\begin{verbatim}
4.66d-31*axionz2ye*faxioncsi*axion_g10**2*T**4.
\end{verbatim}

For a homogeneous composition ($^A_Z\mathrm{X}$), the sum appearing in these equations $\sum_i{Y_i}Z_i^2$, is just equal to $Z/A+Z^2/A$ where the first/second term corresponds to the electron/ion contribution. That gives a factor of e.g. 2 for pure $^1$H or 3/2 for pure $^4$He. However, \cite{Friedland13} use for this sum,
\begin{verbatim}
      ye    = zbar * abari
      axionz2ye=z2bar+ye
\end{verbatim}
which translates into $\bar{Z}^2+\bar{Z}/\bar{A}$ (Mads Soerensen priv. communication). The sum would then be calculated incorrectly, e.g. 4+2/4 instead of 6/4 for pure $^4$He, i.e. a factor of 3 difference, that will increase during subsequent burning phases.

\end{document}